\documentclass[12pt,preprint]{aastex}
\newcommand{\mean}[1]{\mbox{$\langle#1\rangle$}} % generic mean for defined qu.

\shorttitle{Bolocam Survey of Perseus}
\shortauthors{Enoch et al.}

\begin{document}


\title{Bolocam Survey for 1.1~mm Dust Continuum Emission in the c2d Legacy Clouds. I. Perseus}


\author{Melissa L. Enoch (1), Kaisa E. Young (2), Jason Glenn (3), Neal J. Evans, II (2), Sunil Golwala (1), Anneila I. Sargent (1), Paul Harvey (2), James Aguirre (3), Alexey Goldin (4), Douglas Haig (5), Tracy L. Huard (6), Andrew Lange (1), Glenn Laurent (3), Phil Maloney (3), Philip Mauskopf (5), Philippe Rossinot (1), and Jack Sayers (1)}
\email{MLE: menoch@astro.caltech.edu, KEY: kaisa@astro.as.utexas.edu, JG: Jason.Glenn@colorado.edu}
\affil{(1) Division of Physics, Mathematics \& Astronomy, California Institute of Technology,
    Pasadena, CA 91125 \\ (2) The University of Texas at Austin, Astronomy Department, 1 University Station C1400, Austin, TX, 78712-0259 \\ (3) Center for Astrophysics and Space Astronomy, 389-UCB, University of Colorado, Boulder, CO 80309 \\ (4) Jet Propulsion Laboratory, California Institute of Technology, 4800 Oak Grove Drive, Pasadena, CA 91109 \\ (5) Physics and Astronomy, Cardiff University, 5, The Parade, P.O. Box 913, Cardiff, CF24 3YB Wales, UK \\ (6) Harvard-Smithsonian Center for Astrophysics, 60 Garden St., Cambridge, MA 02138}


%\altaffiltext{5}{Patron, Alonso's Bar and Grill}


\begin{abstract}

We have completed a 1.1~mm continuum survey of 7.5 deg$^2$ of the Perseus Molecular Cloud using Bolocam at the Caltech Submillimeter Observatory.  This represents the largest millimeter or submillimeter continuum map of Perseus to date.  Our map covers more than 30,000 $31\arcsec$ (FWHM) resolution elements to a $1\sigma$ RMS of 15~mJy/beam.
We detect a total of 122 cores above a $5 \sigma$ point source mass detection limit of $0.18M_{\sun}$, assuming a dust temperature of $T_D=10$~K, 60 of which are new millimeter or submillimeter detections. 
The 1.1~mm mass function is consistent with a broken power law of slope $\alpha_1=1.3~(0.5M_{\sun}<M<2.5M_{\sun})$ and $\alpha_2=2.6~(M>2.5M_{\sun})$, similar to the local initial mass function slope ($\alpha_1=1.6~ M<1M_{\sun}$,  $\alpha_2=2.7~M>1M_{\sun}$).
No more than 5\% of the total cloud mass is contained in discrete 1.1~mm cores, which account for a total mass of $285M_{\sun}$.
We suggest an extinction threshold for millimeter cores of $A_V\sim5$~mag, based on our calculation of the probability of finding a 1.1~mm core as a function of $A_V$. 
Much of the cloud is devoid of compact millimeter emission; despite the significantly greater area covered compared to previous surveys, only $5-10$ of the newly identified sources lie outside previously observed areas.
The two-point correlation function confirms that dense cores in the cloud are highly structured, with significant clustering on scales as large as $2 \times 10^5$~AU.  
Our 1.1~mm emission survey reveals considerably denser, more compact material than maps in other column density tracers such as $^{13}$CO and $A_V$, although the general morphologies are roughly consistent.  
These 1.1~mm results, especially when combined with recently acquired c2d Spitzer Legacy data, will provide a census of dense cores and protostars in Perseus and improve our understanding of the earliest stages of star formation in molecular clouds.

\end{abstract}


%% subject header.  Objects should be in the appropriate "individual"
%% headers (e.g. quasars: individual, stars: individual, etc.) with the
%% additional provision that the total number of headers, including each
%% individual object, not exceed six.  The \objectname{} macro, and its
%% alias \object{}, is used to mark each object.  The macro takes the object
%% name as its primary argument.  This name will appear in the paper
%% and serve as the link's anchor in the electronic edition if the name
%% is recognized by the data centers.  The macro also takes an optional
%% argument in parentheses in cases where the data center identification
%% differs from what is to be printed in the paper.


%\keywords{globular clusters: general ---
%globular clusters: individual(\objectname{NGC 6397},
%\object{NGC 6624}, \objectname[M 15]{NGC 7078},
%\object[Cl 1938-341]{Terzan 8})}


\section{Introduction}

Observations of the earliest stages of molecular core collapse and protostellar formation are extremely important pieces in the puzzle of low mass star formation, not only illuminating the transition from dense cloud cores to later protostellar phases, but also providing clues about initial conditions and a crucial link between theoretical and empirical scenarios.
Prestellar cores have no internal luminosity source and are therefore very cold ($5-20$~K), with spectral energy distributions (SEDs) that peak at submillimeter or millimeter wavelengths.  
At very early times in an embryonic protostar's life, generally referred to as the Class~0 phase \citep{and93}, it is deeply embedded in an obscuring envelope of gas and dust.  Most of the continuum emission from the hot young star is absorbed and re-radiated by the cool ($10-30$~K) dust envelope at far-infrared (far-IR) to millimeter wavelengths.

We follow \citet{ge00} and distinguish prestellar cores, which are expected to eventually form a star or stellar system, from starless cores, which are dense cores without IRAS sources that may or may not collapse in the future, by the presence of submillimeter or millimeter emission.  \citet{ge00} found that strong submillimeter emission correlates well with collapse signatures and the presence of NH$_3$, making these sources likely to be prestellar.
In this paper, we refer to a 1.1~mm core as any discrete source of 1.1~mm emission, a prestellar core as any 1.1~mm core lacking far-IR emission, and a protostar as any protostellar object with a substantial envelope (Class 0/I; see \citet{lada87,and93} for a description of classifications). 

The millimeter wavelength range is an ideal region in which to study such objects for several reasons.  At these long wavelengths thermal emission from dust becomes optically thin, and the observed flux density traces the total mass of dust in cores and envelopes.  Because they lack an internal luminosity source, prestellar cores are generally invisible at shorter wavelengths.  Additionally, flux density measurements in the millimeter contribute important information to the SED, helping to clarify the evolutionary state of a given object.
Although deeply embedded objects have remained relatively elusive due to the difficulty of observing at submillimeter and millimeter wavelengths, continually improving technology and instrumentation are now making sensitive observations in this regime possible. 

Large format bolometer arrays like SCUBA \citep{holl99}, SHARC II \citep{dow03} and MAMBO \citep{krey98} have made it feasible to scan relatively large fields at continuum submillimeter and millimeter wavelengths, while interferometers such as OVRO \citep{wood98}, BIMA \citep{welch96}, PdBI \citep{guill92} and the SMA \citep{ho04} begin to resolve the details of individual sources.
Nevertheless, time constraints have, for the most part, prohibited coverage of entire molecular clouds with current millimeter continuum instruments. 
Large, complete, high resolution surveys of entire star forming regions are necessary to minimize bias and systematic effects, improve statistics, and develop a clear framework within which to interpret the many observations of individual objects that are now available.
Bolocam, a 144-element bolometer array designed for mapping large fields at millimeter wavelengths \citep{glenn03}, is well suited to the kind of large scale surveys now needed. 

Nearby molecular clouds such as Perseus, Ophiuchus and Serpens, where there is considerable evidence of ongoing star formation \citep[e.g.][]{evans99}, provide the best opportunity to observe stars in the earliest stages of their formation.
Perseus is located in the larger Taurus-Auriga-Perseus dark cloud complex at about $3^{h}30^{m}$, $+31\degr$, and extending approximately $7\degr$ in Right Ascension (RA) and $3\degr$ in Declination (Dec).
The Perseus cloud is often cited as an intermediate case between the low-mass, quiescent Taurus and turbulent, high-mass Orion star formation regions \citep[e.g.][]{ladd93,ladd94}, making it perhaps an ideal environment for studying ``typical'' low mass star formation.  
  Two young clusters lie in Perseus:  IC 348 is a young infrared cluster of age about $2$Myr containing several hundred members of total mass about $160 ~M_{\sun}$ \citep{luhm03}.  NGC 1333 is a very young ($<1$Myr) highly obscured cluster with about 150 stars totaling $\sim 79 ~M_{\sun}$ \citep{lada96,wilk04}, and evidence of ongoing star formation.
Perseus contains several Lynds and Barnard dark clouds, including B5 at the eastern end, B1, and L1455 and L1448 at the western end. 
A number of dense ammonia (NH$_3$) cores have also been identified by \citet{ladd94}.

Recent estimates of the distance to Perseus range from 220~pc to 350~pc \citep[e.g][]{cernis90,hj83}.  
Larger values ($300-350$~pc) are often adopted based on the Perseus OB2 association, which has a fairly well established distance of $\sim330$ pc from Hipparcos parallax and reddening measurements of its members \citep[e.g.][]{bb64,dez99}. 
There is evidence, however, that Per OB2 may lie well behind the complex of dark clouds in which we are interested, and are probably at a distance closer to 250~pc \citep{cernis93, bel02}.  
On the other hand, a single distance for the whole cloud might not be appropriate.
Early CO observations \citep{sarg79} indicated a smoothly varying LSR velocity gradient across the cloud from $v=+3$km~s$^{-1}$ at the western end to $v=+10$km~s$^{-1}$ at the eastern end.  
Given this gradient in velocity, there may also be a distance gradient across the cloud.  Extinction studies of several different regions point to an increase in distance from 220~pc to 260~pc moving from west to east \citep{cernis90, cernis93, cs03}.  

Another possibility is the superposition of two or more clouds.  Based on star counts, \citet{cern85} suggest two dust layers at $d\sim200$~pc and $d\sim300$~pc.
New data compiled by the COMPLETE team also indicate that Perseus may actually be composed of several separate clouds at different distances, projected together on the sky \citep{ridge05}. 
We acknowledge that the structure and dynamics of the Perseus cloud are complicated; it is possible, perhaps likely, that it will ultimately be shown that Perseus is in fact a superposition of a number of smaller clouds.  For the purpose of this work, however, we adopt a distance of 250~pc for the entire area surveyed, based on the most recent extinction studies and parallax measurements of IC~348 members \citep{cernis93, cs03, bel02}. 

Perseus has been fully mapped in CO isotopes tracing densities up to a few thousand particles per cubic centimeter \citep{pad99,ridge05}.  Most previous
submillimeter and millimeter continuum and molecular line mapping of higher-density gas tracers, however, has been confined to the dense cluster region of NGC~1333 \citep[e.g.][]{SK01}, or focused on bright IRAS sources \citep[e.g.][]{ladd94} or energetic outflow sources \citep[e.g.][]{froe03,MW02,bars98}, with the notable exception of the recent SCUBA survey by \citet{hatch05}.  
Conclusions about the cloud based on such existing observations may be problematic.
The gas traced by CO observations has relatively low density and therefore is not necessarily representative of current star formation activity, while isolated small-scale studies may be providing a biased picture of the cloud.

The observations presented here have the distinct advantage that they overlap entirely with the Perseus fields observed with IRAC and MIPS as part of the Spitzer Legacy Project, ``From Molecular Cloud Cores to Planet Forming Disks'' \citep[``Cores to Disks'' or c2d;][]{evans03}.  
The c2d project has mapped five large star forming clouds with the IRAC ($\lambda=3.6-8.0 ~\micron$) and MIPS ($\lambda=24-160 ~\micron$) instruments on Spitzer.  
While millimeter observations are essential to trace core and envelope mass and directly observe prestellar cores, IR observations are necessary to characterize embedded protostars.  Additionally, IR measurements are more sensitive to temperature than to total mass, making them an excellent complement to millimeter observations.

We have completed millimeter continuum observations of the entire Perseus cloud with Bolocam at the Caltech Submillimeter Observatory (CSO). 
This map, observed at $\lambda=1.1$mm during 2003, January-February, is the first unbiased, flux-limited survey of cores and protostars in Perseus at millimeter wavelengths. 
As part of the same project, the remaining two large \textbf{c2d} clouds accessible from Mauna Kea, Ophiuchus and Serpens, also have coordinated large scale Bolocam 1.1~mm and Spitzer observations (Young et al. 2005, in preparation; Glenn et al. 2005, in preparation). 
When joined with the Perseus data, this combined sample will provide a unique basis for comparing star forming properties in varying environments, without the systematic difficulties introduced by observations from different instruments or variable coverage. 
Detailed SEDs for this complete sample of objects will be made possible through the combination of Spitzer IR and Bolocam 1.1~mm fluxes, as well as 850~$\micron$ fluxes when available, allowing the construction of a more quantitative evolutionary sequence than currently available, and calculation of statistical lifetimes.

Here we present the results of our Bolocam survey of Perseus.  In Section~\ref{obssect} we describe the observations, and in Section~\ref{redsect} we describe in detail the reduction techniques used for the Perseus Bolocam data, including an iterative mapping scheme we have developed to restore source brightness lost during sky subtraction.  Results are presented in Section~\ref{ressect}, including source flux, size, and mass statistics and a comparison of the cloud appearance at 1.1~mm to other column density tracers.  In Section~\ref{discsect} we discuss the completeness limits of the survey and the mass versus size distribution, the 1.1~mm Perseus mass function, and the clustering characteristics of the cloud.  We end with a summary and plans for future work in Section~\ref{sumsect}.

\section{Observations}\label{obssect}

%% In a manner similar to \objectname authors can provide links to dataset
%% hosted at participating data centers via the \dataset{} command.  The
%% second curly bracket argument is printed in the text while the first
%% parentheses argument serves as the valid data set identifier.  Large
%% lists of data set are best provided in a table (see Table 3 for an example).
%% Valid data set identifiers should be obtained from the data center that
%% is currently hosting the data.

Continuum observations at 1.1~mm were made with Bolocam\footnote{\url{http://www.cso.caltech.edu/bolocam}} at the Caltech Submillimeter Observatory (CSO) on Mauna Kea, Hawaii during 2003, January 28-February 15.  
Bolocam is a 144-element bolometer array that operates at $\lambda =$ 1.1, 1.4, and 2.1~mm. 
The instrument consists of a monolithic wafer of silicon nitride micromesh AC-biased bolometers, cooled to 260 mK.  During the Perseus observations 81 of 144 bolometers were operational. The field of view (FOV) is $7\farcm5$ and the beams are well approximated by a Gaussian with a FWHM of 31$\arcsec$ at $\lambda=1.1$mm.  The focal plane is flat over the FOV due to a cold field lens coupling the telescope to the array \citep{glenn98,glenn03,haig04}.
All observations of Perseus were completed in the 1.1~mm mode, which has a bandwidth of 45 GHz and band center at 268~GHz.  

\begin{figure}
\plotone{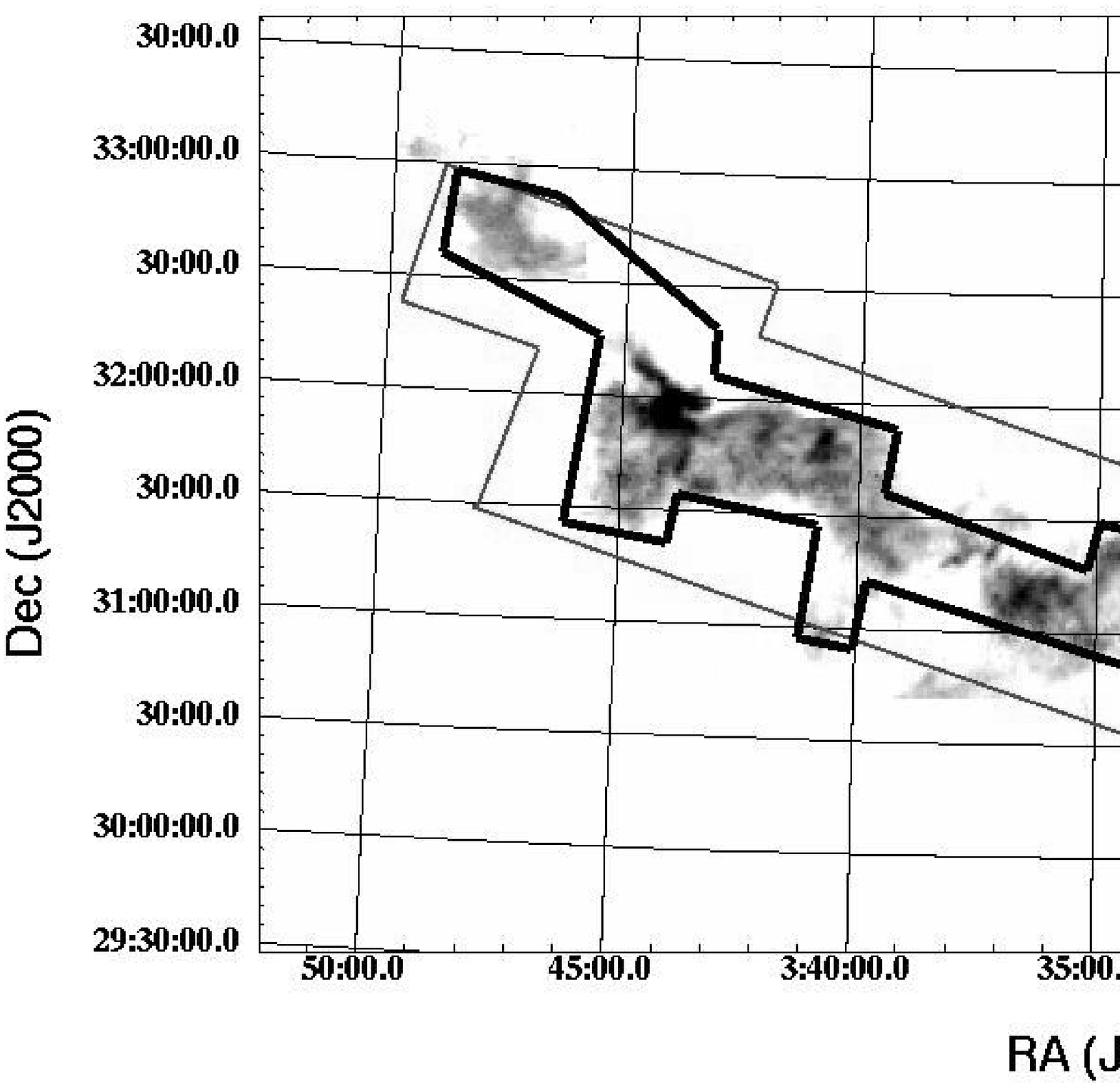}
\caption{Bolocam 1.1~mm (thin line) and Spitzer c2d IRAC (thick line) coverage of Perseus overlaid on a $^{13}$CO integrated intensity map from \citet{pad99}.  The area observed by IRAC was chosen based on the $^{13}$CO intensity, and corresponds approximately to $A_V \ge 2$~mag \citep{evans03}.  The Bolocam observations were designed to cover the same region.  The MIPS data cover a somewhat larger area.  \label{c2dfig}}
\end{figure}

The integrated intensity $^{13}$CO map of \citet{pad99} was used to define the area of the Bolocam observations, as shown in Figure~\ref{c2dfig}.  The chosen boundaries correspond to an approximate extinction limit of $A_V \sim2$~mag \citep{evans03}, and were designed to overlap entirely with the area to be observed with Spitzer as part of the c2d Legacy project.
Maps were made in raster scan mode in sets of three scans offset by $-11\arcsec$, $0\arcsec$, $+11\arcsec$, with a scan speed of 60 $\arcsec$~s$^{-1}$ and a subscan spacing of $162\arcsec$.
A subscan is defined as one pass of the array across the field, and a scan as a set of consecutive subscans that cover the field entirely.  
Offset scans are necessary to obtain a fully sampled map because the array elements are separated on the sky by $1.5 f \lambda$, whereas $0.5 f \lambda$ corresponds to Nyquist sampling.
Simulations indicated that $11\arcsec$ offsets were optimal for obtaining the best coverage perpendicular to the scan direction.

The total area was divided into three large rectangles for the most efficient scanning.
On each of 19 nights Perseus was observed for approximately two hours before and after transit, with a total of 12 scans of each section completed.  A number of subscans have been omitted from the final map due to bad weather, manifested as very large sky noise, or temporarily poor bolometer performance.
The total observing time was about 40 hours for the 7.5 square degree region, with 31 hours actually spent integrating on-source, or an observing efficiency of 75\%.  This corresponds to a mapping speed for Perseus of 3~arcmin$^2$mJy$^{-2}$hr$^{-1}$. 
No chopping was done for any of the observations, thereby retaining, in principle, sensitivity to large-scale structure up to the angular size of the array ($7\farcm5$).  Chopping is not required because the bolometers are AC-biased, which elevates the signal band above the atmospheric and instrumental $1/f$ noise.  Demodulation brings the signal down to near-DC, with the signal band determined by the scan speed and beam size.

One night at the beginning of the run was devoted to pointing and calibration observations only, in order to set the focus and pointing constants for the run.  
Each night secondary calibrators, including the bright Class~0 object NGC~1333-IRAS4A, were observed approximately every two hours, with calibrator sources from many areas of the sky used to derive a calibration curve for the entire run (see Section~\ref{mapsect}).
All calibrator observations were taken at the same scanning speed as science fields (60$\arcsec$~s$^{-1}$), and most were small $4\arcmin \times4\arcmin$ scan maps. 
At least one primary calibrator, usually Mars, was observed nightly, and six large $10\arcmin\times 10\arcmin$ beam maps made over the 3 week run.  These beam maps are used to define the distortion corrections and the beam shape, which is found to be very gaussian. 
Nightly sky dips were used to measure the sky and telescope optical loading (the quiescent optical power received by the bolometers from the sky, telescope, and dewar). 
Weather was mostly clear for the run, with an average zenith 225 GHz tau of $\tau \sim 0.07$, ranging between $0.05$ and $0.09$.

%% In this section, we use  the \subsection command to set off
%% a subsection.  \footnote is used to insert a footnote to the text.


%% This section also includes several of the displayed math environments
%% mentioned in the Author Guide.

\section{Data Reduction}\label{redsect}

Given that these observations utilize a new instrument, we describe the data reduction process for Bolocam data in general, as well as reduction techniques specific to the Perseus data, in some detail. 
Preliminary data reduction was accomplished using a reduction pipeline written by the Bolocam instrument team, as described below and in \citet{laurent05}.  A number of problems specific to the bright sources and very long subscans of more than a few degrees observed for this project required the development of additional reduction techniques, in particular the iterative mapping routine described in Section~\ref{itmapsect}.

Initial steps in the pipeline include calculating the pointing model as a function of azimuth and elevation, and calculating the RA/Dec of each bolometer for every time sample. 
The bulk of the data reduction effort goes into removing the (considerable) sky noise.
Subsequently, bright pixel spikes from cosmic rays are flagged, and the power spectral density (PSD) of each subscan is calculated.  The $N_{bolo}$ bolometer timestreams are converted into a 2D pixel map using pixel offsets, subscan PSDs, and calibration constants.  Finally, an iterative mapping scheme developed for this purpose is used to recover flux density lost in the sky subtraction process.  

\subsection{Pointing}\label{pointsect}

  All pointing observations made during the 2003, January  observing run, including secondary calibrators as well as additional variable sources such as quasars, are used to calculate precise pointing corrections.  
The resulting pointing model serves to refine the recorded telescope pointing position. 
The RMS of the global pointing model for 2003, January as determined from the positions of known galaxies in the Lockman Hole is $9.1\arcsec$ \citep{laurent05}.  
The pointing model is somewhat better in Perseus, likely due at least in part to the proximity of many pointing sources to the Galactic plane.  

We compared our positions for known sources to the literature, using submillimeter, millimeter, or radio positions when possible, and otherwise IRAS positions.  
There was a small systematic offset in RA ($\delta \mathrm{RA} = -3\farcs 2 \pm1\farcs 3$), but not large enough to warrant a correction to positions, and no systematic offset in Dec ($\delta \mathrm{Dec} =-0\farcs 5 \pm 1\farcs 5$).
We find a $1\sigma$ RMS compared to previous positions of $7\arcsec$, independent of azimuth or elevation.  
Given that this dispersion includes potentially large uncertainties in the literature positions, as well as possible physical offsets between IR and 1.1~mm sources, we conservatively estimate the overall accuracy for point source positions in Perseus to be $7\arcsec$.  
In addition to uncertainties in source positions, pointing errors increase the effective beam size, causing sources to be blurred in the coadded map and affecting the measured size and peak flux density, but not the total flux density.

Position offsets of the various bolometers in the array from the telescope pointing center, as well as optical distortions -- collectively termed pixel offsets -- are measured using fully sampled observations of planets. 
Distortion corrections account for distortion in the optics, due primarily to an off-axis ellipsoidal mirror, and secondarily to imperfect optical alignments.  Corrections are typically of order $2\arcsec-3\arcsec$.
Pixel offsets of each bolometer are used to compute the RA and Dec value of every time sample in a subscan, and later used to convert bolometer timestreams into a 2D pixel map.

\subsection{Removal of Sky Noise}\label{cleansect}

The sensitivity of a given Bolocam observation is determined by the intrinsic sensitivity of and optical loading on the bolometers, the integration time, and the success of sky noise subtraction.  The most important reduction step is the removal of sky noise, or cleaning.  On scales comparable to or larger than the beam, the instrument is limited by $1/f$ noise, which is primarily atmospheric but also instrumental in nature.  Sky noise originates as fluctuations in the brightness temperature, or column density, of the atmosphere and dominates most astronomical sources at $\lambda=1.1$mm.  Because chopping is not done, this noise is present in the bolometer timestreams before cleaning.  

The bolometer beams overlap almost completely in the near-field where the sky noise originates, but do not overlap in the far-field where astronomical signals originate.  Therefore, to first order the sky noise is identical for each bolometer and a sky template can be constructed quite simply by taking an average or median across the bolometer array at each point in time.  This average cleaning method is appealingly simple, but does not deal well with multiple correlated $1/f$ noises with different correlation coefficients, or with $1/f$ noises that are correlated on spatial scales smaller than the array.  The latter might arise if the beams do not overlap completely at the height of the sky noise.  

In principle, the correlated noise can be removed because it is correlated in time, whereas the astronomical signal is correlated in space across the array.  
A more sophisticated approach that addresses this issue is Principal Component Analysis (PCA) cleaning (see~\citet{laurent05} and references therein for a discussion of PCA cleaning).  
In a PCA analysis the raw timestreams are projected along eigenvectors, bringing out common modes, or principal components, in the data.  Patterns common to all bolometers correspond to sky noise, so subtracting such common modes from the data is an efficient sky subtraction technique.
Removing the first principal component is nearly equivalent to performing average subtraction. 

Any number of components can be subtracted from the data, each removing progressively less correlated $1/f$ noise.  The actual reduction in the RMS noise depends on the initial sky noise present, but typically removing 3 PCA components reduces the overall noise by $10-30$\% compared to average cleaning.  
Although removing more components will reduce the noise further, the disadvantage of PCA cleaning is that higher components tend to remove source flux density (most of which can be recovered, see Section~\ref{itmapsect}).  Tests performed on observations of Serpens with both compact and extended sources indicate that removing between 1 and 5 components is most effective at eliminating stripes from $1/f$ noise while retaining source flux density and structure (Glenn et al. 2005, in preparation).

\subsection{Mapping and Calibration}\label{mapsect}

To make a 2D pixel map from the bolometer timestreams, the pointing model and empirically derived pixel offsets are used to project each bolometer time sample onto an RA/Dec grid of pixel values. 
Timestreams are coadded using a weighted (by the inverse of the PSD) average.  
We bin the map at a resolution equal to 1/3 of the true instrumental resolution, or 10$\arcsec$/pixel, which gives sufficient hits per pixel for significant statistics without degrading the resolution.  Given the nature of the instrument and observations, a single pixel in the map will contain data from many bolometers and many scans.  
We refer to the coverage map as an image of the number of hits per pixel, or seconds per pixel, in the map.  The coverage is dependent on pixel size, scan strategy, the number of bolometers, and the number of scans in the map.  The average coverage for Perseus is about 500 hits~pix$^{-1}$ (or 10 s~pix$^{-1}$), varying by 30\% across the map and leading to $\sim15$\% variation in the RMS noise. 

To maximize the signal to noise of, and thus our chance of detecting, point sources, we optimally filter the map.  Because the signal from a point source lies in a limited frequency band, we can use an optimal (Wiener) filter to attenuate $1/f$ noise at low frequencies, as well as high frequency noise above the signal band.  Attenuating the $1/f$ noise reduces the overall RMS \textit{per pixel} by $\sim \sqrt{3}$ (making the RMS/pixel $\sim$ RMS/beam), maximizing the probability of detection for faint point-like sources. 
The optimal filter $g(q)$ is given by:
\begin{equation}
g(q) = \frac{s^{*}(q)/J(q)}{\int \left| s(q) \right|^2/J(q) d^2q}
\end{equation}
where $q$ is the spatial wavenumber, $J(q)$ is the azimuthally averaged PSD, $s(q)$ is the Fourier transform of the beam, and $g(q)$ is normalized so that the peak brightness of point sources is preserved.  
Note that the optimally filtered map is used for source detection only; all maps shown are unfiltered.

The flux calibration factor at any given time depends on the bolometer DC resistance, which changes with atmospheric attenuation and bolometer optical loading.  Therefore, the calibration cannot be applied as a single factor to the final map, but must be calculated as a function of time for each subscan based on the average DC resistance of that subscan.  
This is a powerful calibration method because it makes real-time corrections for the atmospheric attenuation and bolometer operating point using the bolometer optical loading, effectively providing flux calibration on the timescale of minutes. 

In practice, the calibration (in mV/Jy) as a function of the DC voltage is determined using a second-order polynomial fit to observations of Mars, Saturn, and a number of secondary calibrator sources (see \citet{laurent05} for an expanded discussion).
All calibrators with reliable known flux densities (one observed every 1-2 hours per night throughout the run) are used to define the calibration curve, not only those observed during a given science observation.  This enables a comprehensive calibration curve over a range of atmospheric optical depths.  Reference flux densities are from the JCMT planetary flux website\footnote{http://jach/hawaii.edu/JACpublic/JCMT/software/bin/fluxes.pl} and \citet{sand94}, corrected for the Bolocam bandpass.
The effects of non-linearity due to optical loading from the sky and the finite beam size are also accounted for, using large beam maps of Mars and assuming a uniform disk model. 
Measured flux densities are from a gaussian fit after application of a point source filter. 

The resulting peak-normalized map can be divided by the beam area to obtain a surface brightness-normalized map.   The integral of a point source in the surface brightness map returns the total flux of the point source, thus this map is suitable for photometry.  The absolute calibration uncertainty, derived from the deviation of measured calibrator values from the quadratic fit, is 9.7\% ($1 \sigma$; \citealt{laurent05}).

\subsection{Iterative Mapping}\label{itmapsect}

\subsubsection{Method}

Although they utilize different methods, both average and PCA cleaning contain a step that essentially removes the mean of each bolometer subscan (see Section~\ref{cleansect}).  This step is necessary to eliminate sky noise, but when there is a bright source in a subscan it biases the mean.  Consequently, sky subtraction introduces negative lobes around bright sources, which are asymmetric in the scan direction.  Furthermore, subscans containing sources tend to be under-weighted in the coadded map because the source brightness contaminates the integrated PSD, causing a decrease in the weighting factor. 
Both effects tend to suppress source flux density and are mildly dependent on the brightness and structure of the source.  
There is an additional effect due to PCA cleaning that non-linearly removes the flux density of bright sources as more PCA components are removed.  
To correct for diminished source brightness and negative artifacts introduced by the above effects, we have developed an iterative mapping code that robustly restores lost flux density and structure to the map.

The iterative mapping algorithm we have implemented iteratively subtracts a source model from the real data (somewhat similar to CLEAN \citep{hog74,schwarz78}, but working in the image plane).  The following is a more robust method than using, for example, a source model comprised of the sum of many gaussians, because many of our sources are likely to be extended and non-gaussian. 
For the following, $j$ refers to bolometer number ($j=1-81$), and $i$ refers to the iteration number ($i=0-N$, where the zeroth iteration $i=0$ indicates raw or cleaned data before any source model has been subtracted).  
We begin with the raw timestream data, $t_{j,i=0}$.  These data are sent through the cleaning and mapping process to produce the zeroth iteration map, $M_{i=0}$, which contains negative artifacts and is missing some fraction of the flux density of each source.  Next, a cut is applied at $+N_{\sigma} \sigma$ to $M_{i=0}$, removing any negative pixels as well as most of the noise (pixels with values $\le +N_{\sigma}$ times the RMS noise are set to zero).  
We now have a source model map, $M'_i$, our current best guess of the true source flux density.  

From the source model $M'_i$ a model timestream $t'_{j,i}$ is generated for each bolometer and subtracted from the raw timestreams, $dt_{j,i} = t_{j,0} - t'_{j,i}$.  The difference timestreams $dt_{j,i}$ contain residual source flux density that was missing from the original map.  When the difference timestreams are subsequently cleaned and mapped to produce a residual map $dM_i$, there is much less contamination of the sky template by source brightness, so the negative artifacts are greatly reduced.  A threshold cut at $+N_{\sigma} \sigma$ is again applied to $dM_i$ producing a residual source model $dM'_i$, which is added to the original source model to create a new source model for the next iteration ($M'_{i+1} = M'_i + dM'_i$).  This process is iterated until there is no remaining residual source flux density in the difference timestream.  After $N$ iterations, the final residual map $dM_N$, containing only noise and any source flux below the threshold cutoff, is added to the last source map to create the final map ($M_N = M'_{N-1} + dM_N$).

\subsubsection{Performance}

The performance of iterative mapping depends on two parameters:  the number of PCA components removed during cleaning ($N_{PCA}$), and the RMS threshold cut ($N_{\sigma}$) used to make the source model.  While iterative mapping greatly improves the appearance and photometric reliability of the map, our chosen values of $N_{PCA}$ and $N_{\sigma}$ will introduce some systematic uncertainties into the final photometry.  Using Monte Carlo simulations, which consisted of inserting simulated sources into the real map before the iterative mapping process, we found that $N_{PCA}=3$ and $N_{\sigma}=2$ produce the best combination of accurate final photometry (recovery of lost source flux density) and reduced noise compared to average cleaning (by $\sim 20-30 \%$).  Using a value of $N_{\sigma}<2$ tends to introduce noise into the source model, increasing the RMS noise and raising the background level of the map; increasing $N_{\sigma}$ leaves source flux density out of the source model, leaving some negative artifacts and making it difficult to recover extended source features.  Decreasing $N_{PCA}$ leaves significantly more $1/f$ noise in the map and increases the overall RMS noise;  if we increase $N_{PCA}$ much above 3 it again becomes difficult to fully correct for the loss of source brightness, which is severe and non-linear for $N_{PCA}>5$.  

\begin{figure}
\plotone{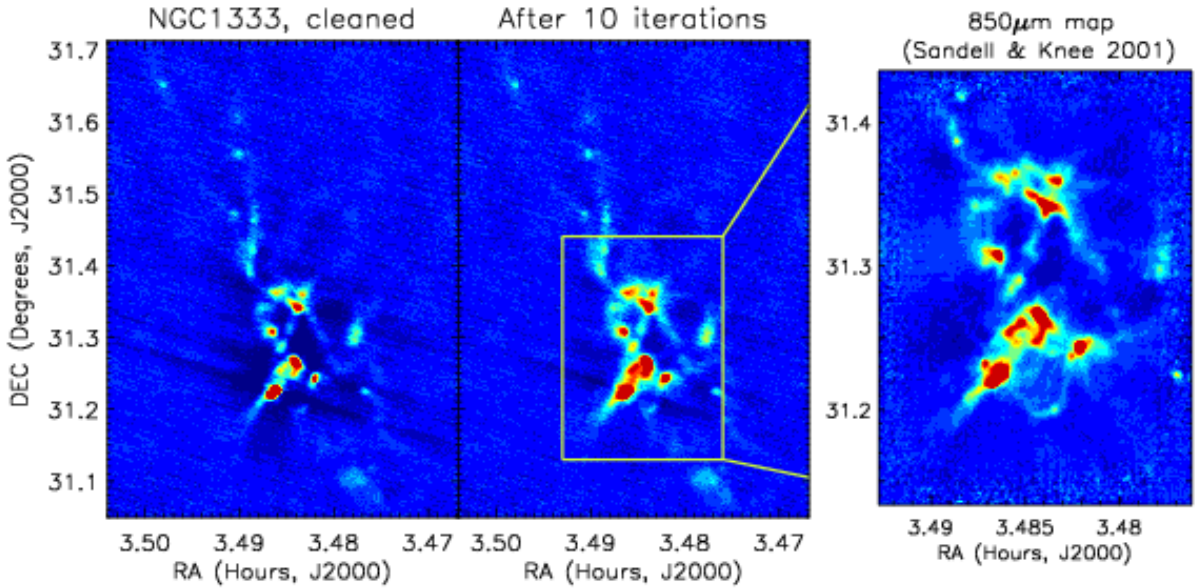}
\caption{NGC~1333, before (left) and after (center) iterative mapping, shown on the same intensity scale.  Dark blue regions are negative lobes introduced by cleaning.  These images illustrate the effectiveness of iterative mapping in reducing such negative artifacts and restoring source flux density lost during PCA cleaning.   After 10 iterations, the brightest source has increased in peak brightness by 14\%.  Although some negative artifacts remain, they are greatly reduced both in extent and intensity (the most negative pixel has decreased in amplitude from $-238$~mJy/beam to $-88$~mJy/beam). No fine-scale structure is lost, and all recovered extended structure is real; for comparison we show the $850~\micron$ map of \citet{SK01}, with a resolution of $14\arcsec$, on approximately the same intensity scale.  
\label{iterfig}}
\end{figure}

Figure~\ref{iterfig} demonstrates the results of iterative mapping on a portion of the Bolocam map containing the crowded region NGC~1333.  The left panel shows NGC~1333 after PCA cleaning only with $N_{PCA}=3$, where the dark blue areas are negative bowls caused by cleaning.  The same region after 10 iterations is shown on the same intensity scale in the center panel.  All sources have increased considerably in brightness (the brightest peak by 14\%), and most of the negative features have been corrected.  
Although some negative artifacts remain, their extent and intensity is greatly reduced; the worst remaining negative pixel in the iterated map is $-88$~mJy/beam, compared to $-238$~mJy/beam in the cleaned map.  
For comparison, the $850~\micron$ map of the central portion of NGC~1333 from \citet{SK01} is also shown on approximately the same intensity scale.  Clearly, the extended structure in the higher resolution ($14\arcsec$) $850~\micron$ map is recovered in the 10 iteration Bolocam map.  The recovered structure is almost certainly real, therefore, and we are not missing any structure at 1.1~mm that is present at $850~\micron$ (although \citet{SK01} note that flux densities for extended emission in their map are unreliable due to similar residual artifacts).

\begin{figure}
%\plotone{f3.eps}
\plotone{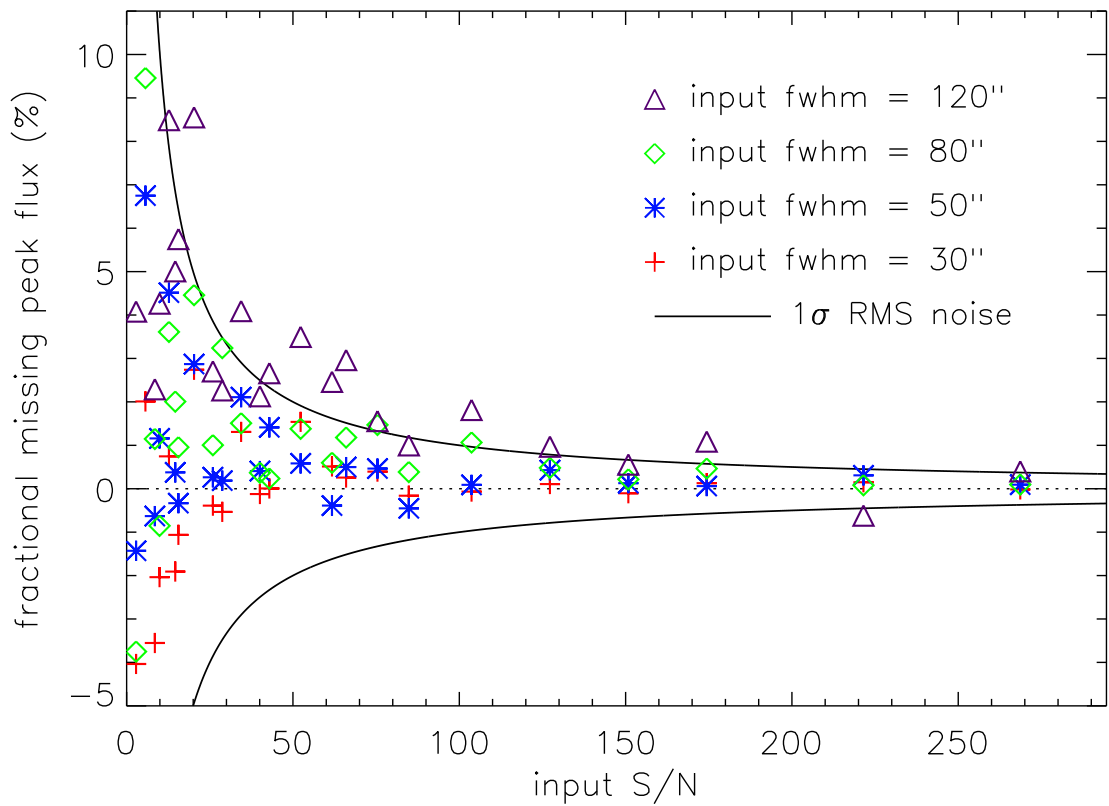}
\caption{Fractional lost peak flux density of simulated sources after 10 iterations, as a function of the input signal to noise (S/N) and FWHM.  Solid lines are the 1 sigma RMS noise as a fraction of the input amplitude, indicating the spread in recovered peak that might be expected from noise alone.  Except for the largest ($120\arcsec$ FWHM) sources, most points lie within the RMS noise, so residual systematic effects from cleaning are not important for measured peak flux densities.  \label{testpeakfig}}
\end{figure}

Monte Carlo tests were done to quantify the effectiveness of the iterative mapping and characterize any systematic errors introduced throughout the cleaning and iterative mapping process, using $N_{PCA}=3$ and $N_{\sigma}=2$.  Simulated gaussian sources of varying amplitudes ($3-300 \sigma$) and sizes ($30\arcsec-250\arcsec$ FWHM) were inserted into an empty region of the map, then cleaned and iterated.  Finally, the resulting peak amplitude, flux density and FWHM were measured.  The fractional missing peak flux density after 10 iterations for a range of test sources is shown as a function of the input peak amplitude in Figure~\ref{testpeakfig}.  Sources of FWHM $30\arcsec$, $50\arcsec$, $80\arcsec$ and $120\arcsec$ are indicated.  The 1$\sigma$ RMS noise per beam is also plotted as a percentage of the input amplitude.  The fact that most points lie beneath this line indicates that the residual bias in measured flux densities after iterative mapping are consistent with the RMS noise.  Only for the largest sources ($120\arcsec$ FWHM, few of which are found in the real map),  is there a residual systematic reduction in flux density that is larger than the RMS.  
The measured FWHM, axis ratio, and position angle are not significantly affected by either cleaning or iterative mapping. 

\begin{figure}
%\plotone{f4.eps}
\plotone{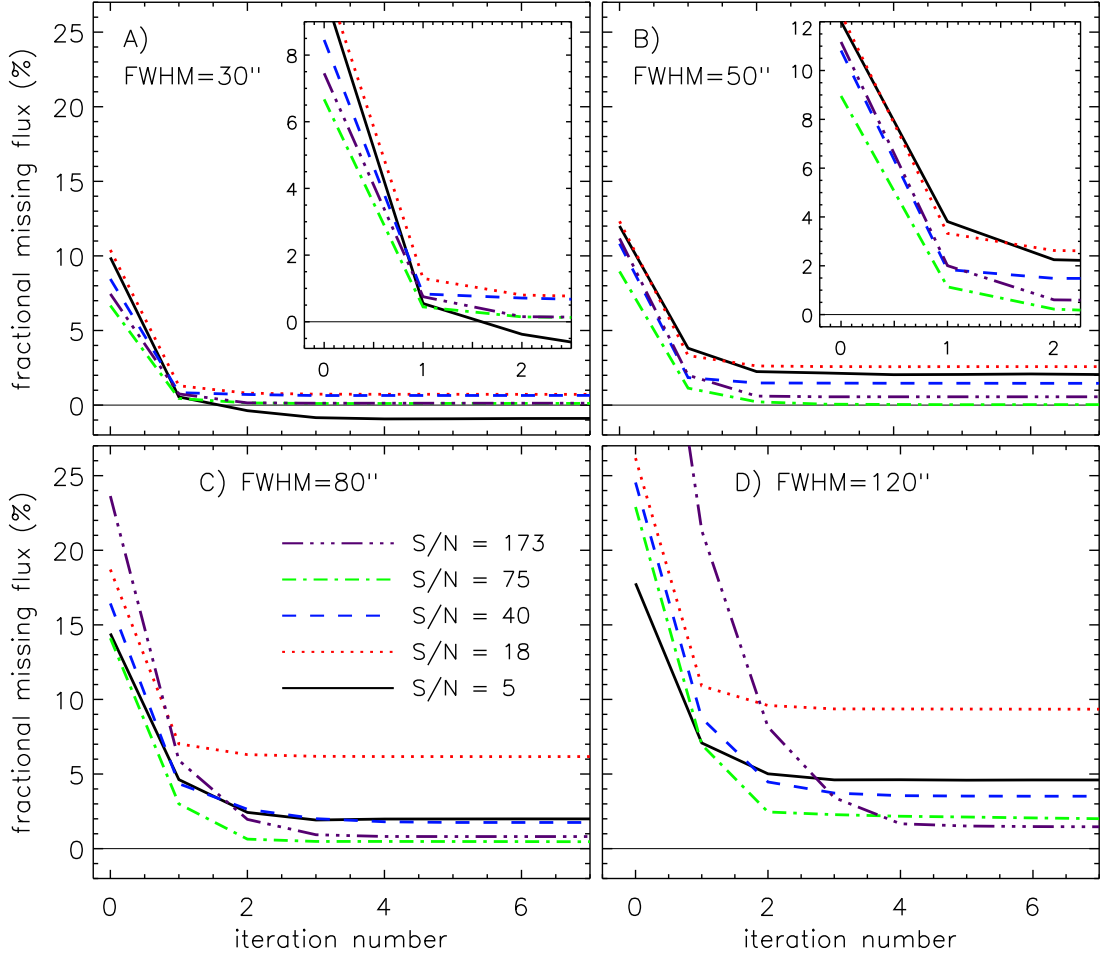}
\caption{Fractional lost integrated flux density in a $40\arcsec$ aperture for simulated sources, as a function of input signal to noise (S/N), FWHM, and iteration number.  Representative source FWHM of $30\arcsec,50\arcsec,80\arcsec,120\arcsec$ are shown, with insets to magnify confused regions.  Larger sources  require more iterations to recover the input flux density, and tend to have the most remaining fractional missing flux density after 10 iterations.  Except for sources with FWHM$\gtrsim 100\arcsec$ the flux has converged by $\sim5$ iterations.  The integrated flux density is usually recovered to within 5\%, and to within 10\% even for large (FWHM$\ge80\arcsec$), faint (S/N$\le20$) sources.  Note that the $1 \sigma$ RMS in a $40\arcsec$ aperture is $\ge10\%$ for sources with S/N$\le18$.  
\label{testfluxfig}}
\end{figure}

Figure~\ref{testfluxfig} shows the fractional lost flux density in a $40\arcsec$ aperture as a function of iteration number for four representative FWHM sizes.  A range of input amplitudes from $S/N=5$ (detection limit) to $S/N=175$ are plotted as different line styles.  Large, bright sources (bottom right panel) are the most affected by cleaning -- more than 50\% of the flux density is removed during PCA cleaning -- but are almost entirely corrected by iterative mapping, with only $\sim2$\% residual missing flux density after 5 iterations.  
For large faint sources $\lesssim 20 \sigma$ (e.g. the dotted line, bottom right), iterative mapping is unable to recover all the flux density to better than $5-10$\%, but note that the $1\sigma$ RMS in a $40\arcsec$ aperture is $\ge 10\%$ for sources $\le 18\sigma$.  
Sources larger than $\sim200\arcsec$ FWHM (not shown) are not fully recovered even after 20 iterations, making this the effective maximum source size detectable.  No sources larger than $150\arcsec$ FWHM are measured in the real map.

We conservatively assign a systematic uncertainty of 5\% to all integrated flux densities.  In almost all cases the measured flux density is lower than the true value, making the uncertainty a bias in that measured values underestimate the true flux densities.  We do not attempt to correct for this residual bias, instead including the 5\% iterative mapping uncertainty in the overall systematic uncertainty of 15\%.

\section{Results}\label{ressect}

\begin{figure}
\plotone{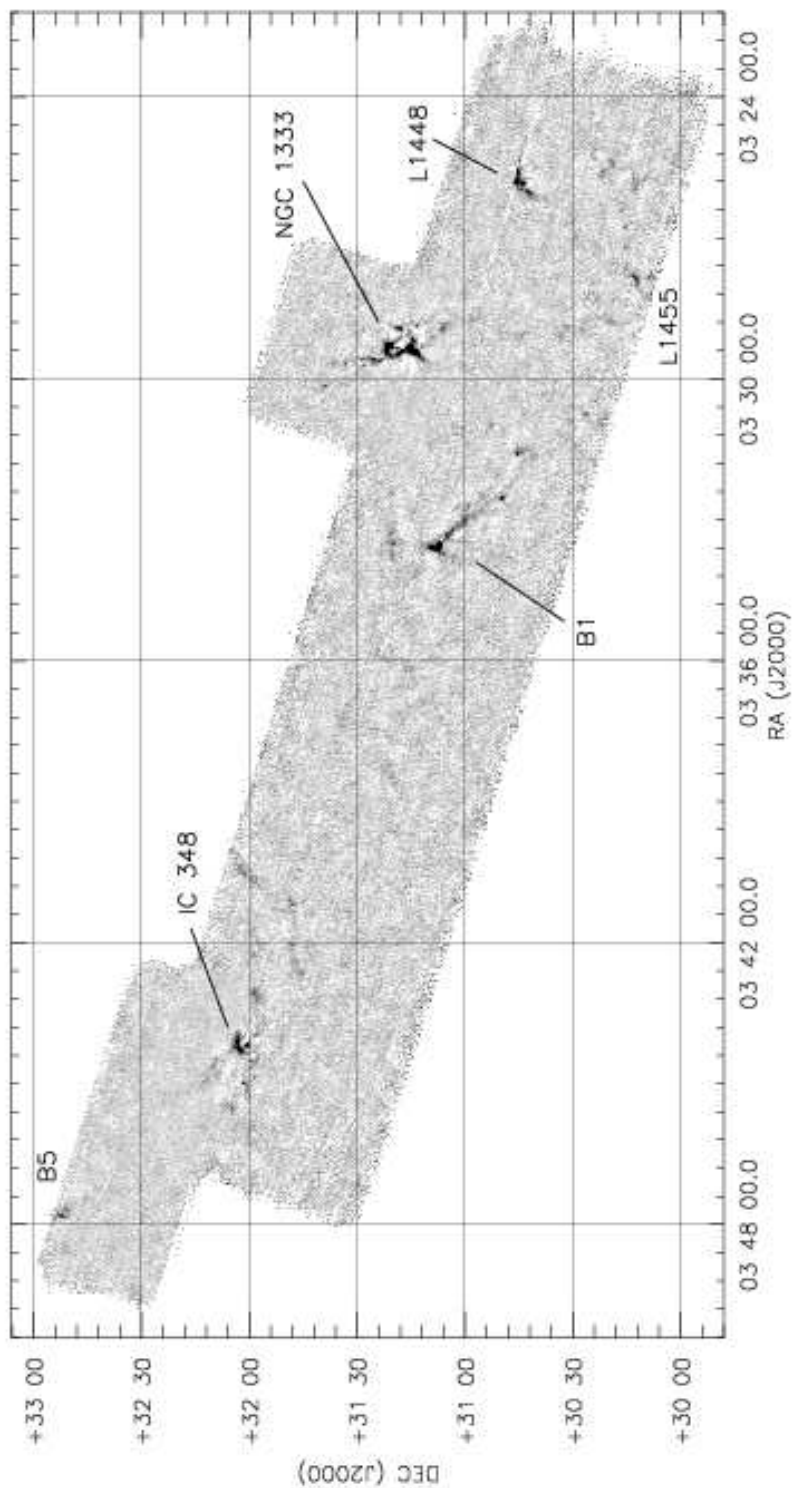}
\caption{Bolocam 1.1~mm map of the Perseus Molecular Cloud (10$\arcsec$/pixel). The $31\arcsec$ resolution map covers 7.5 deg$^2$ (143 pc$^2$ at $d=250$~pc), or 3.4$\times10^4$ resolution elements.  The average $1\sigma$ RMS is 15~mJy/beam, varying by 15\% across the map due to variable coverage.  In this and other figures all maps shown are unfiltered. 
\label{mapfig}}
\end{figure}

The final coadded $10\arcsec$/pixel Perseus map is shown in Figure~\ref{mapfig}.  
The map has a beam size of $31\arcsec$ and a total area of 7.5 square degrees, covering a total of $3.4\times 10^4$ resolution elements. 
The mean RMS is 15 mJy/beam.
Well known regions are marked, including the conspicuous bright cluster NGC~1333.  Note that the infrared cluster IC~348 actually lies slightly to the northeast of the group of 1.1~mm sources indicated as IC~348 here.  

While nearly all previously known sources are seen, with the exception of some crowded regions where blending occurs (e.g. in NGC~1333 and L1448), perhaps the most striking feature of the map is its relative emptiness.  In fact, few new sources are detected far from known cluster regions, and those that are tend to be faint objects near the detection limit. 
There does not appear to be any symmetric extended emission $\gtrsim 3\arcmin$ in the map. 
Although it should in principle be possible to recover symmetric structures up to the array size ($7\farcm5$), our simulations show that sources $\gtrsim 4\arcmin$ are severely affected by cleaning and difficult to fully recover with iterative mapping.

\subsection{Source Identification}\label{idsect}

Discrete sources of 1.1~mm emission, or cores, are identified within the optimally filtered, iterated map using a simple peak finding algorithm.  An optimal filter (as described in Section~\ref{mapsect}) is applied to the iterated map to facilitate the detection of point sources.  This decreases the noise in the map, as well as increasing the peak brightness (and therefore the probability of detection) of sources larger than the beam.  After optimal filtering the noisy edges of the map are removed using a cutoff based on the coverage map.  A cutoff at 20\% of the peak coverage is chosen empirically based on the number of false edge sources detected.  All peaks 5$\sigma$ above the average RMS are flagged as possible sources. 

To make the final cut, candidate sources must also be $5\sigma$ above the local RMS, and have a well defined centroid.  
The centroid is a weighted average position based on the surface brightness within a specified aperture.
The local RMS noise per beam is calculated in small ($\sim 45$ arcmin$^2$) boxes in a noise map, from which sources have been removed using the source model generated during iterative mapping (see Section~\ref{itmapsect}).  The average RMS is $15$~mJy/beam, varying across the map by 15\%.  Most of the variation in RMS is due to 30\% variations in observational coverage rather than to a change in the calibration, which is very stable. 
Peaks separated by more than one beam size from a previously identified source centroid are considered to be distinct sources.  A few false sources at the edge of the map were not automatically cut and had to be removed by visual inspection.  

\begin{figure}
\plotone{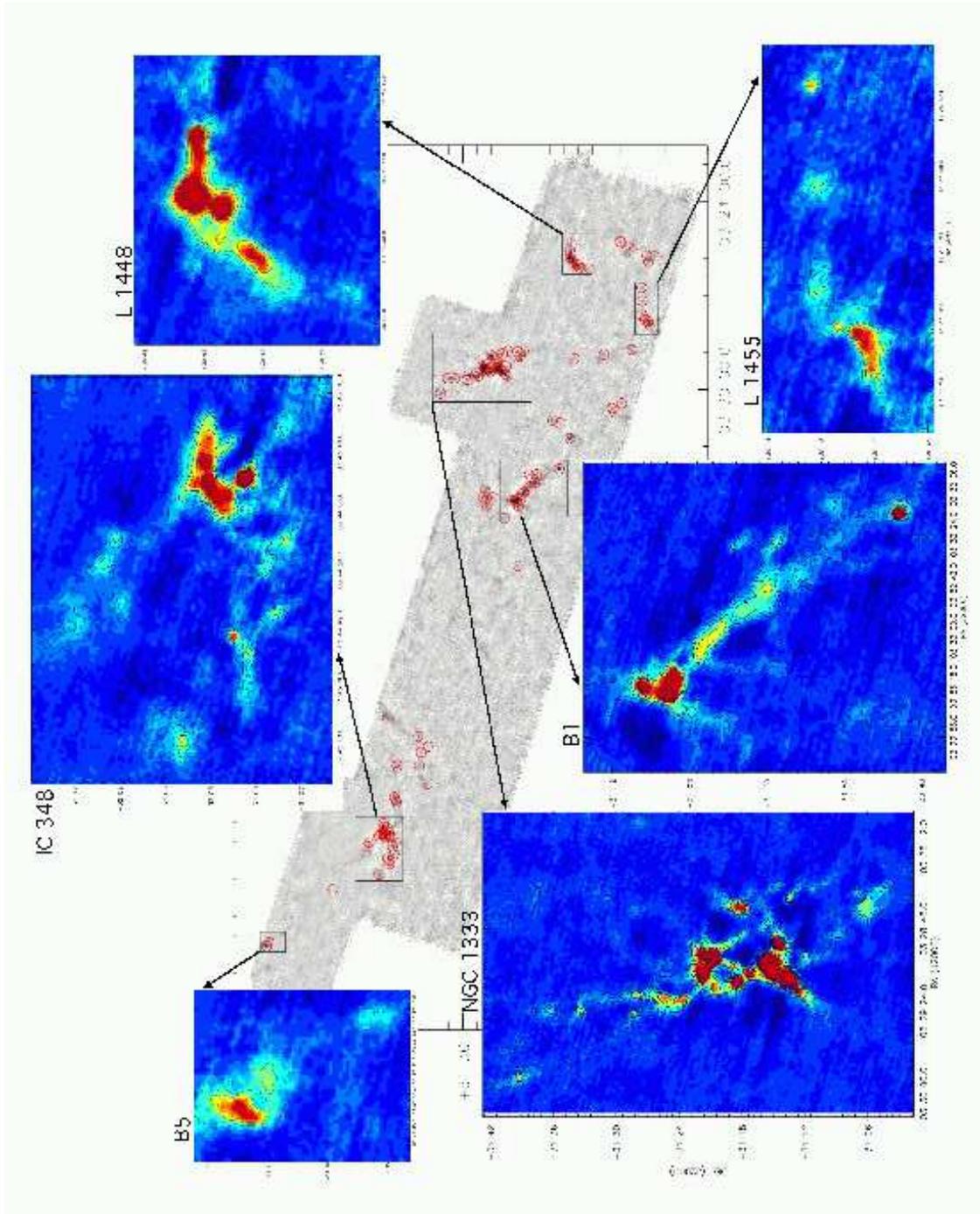}
\caption{Bolocam 1.1~mm map of Perseus, with the 122 1.1~mm sources detected above $5 \sigma$ indicated by small circles.  The RMS varies very little across the map (15\%), so the apparent lack of sources over large regions of the image is real. Despite the greater area surveyed compared to previous work, few new sources are found; much of the cloud is devoid of 1.1~mm emission at this sensitivity.  Regions of high source density are magnified for clarity.  \label{sourcefig}}
\end{figure}

\begin{figure}
\plotone{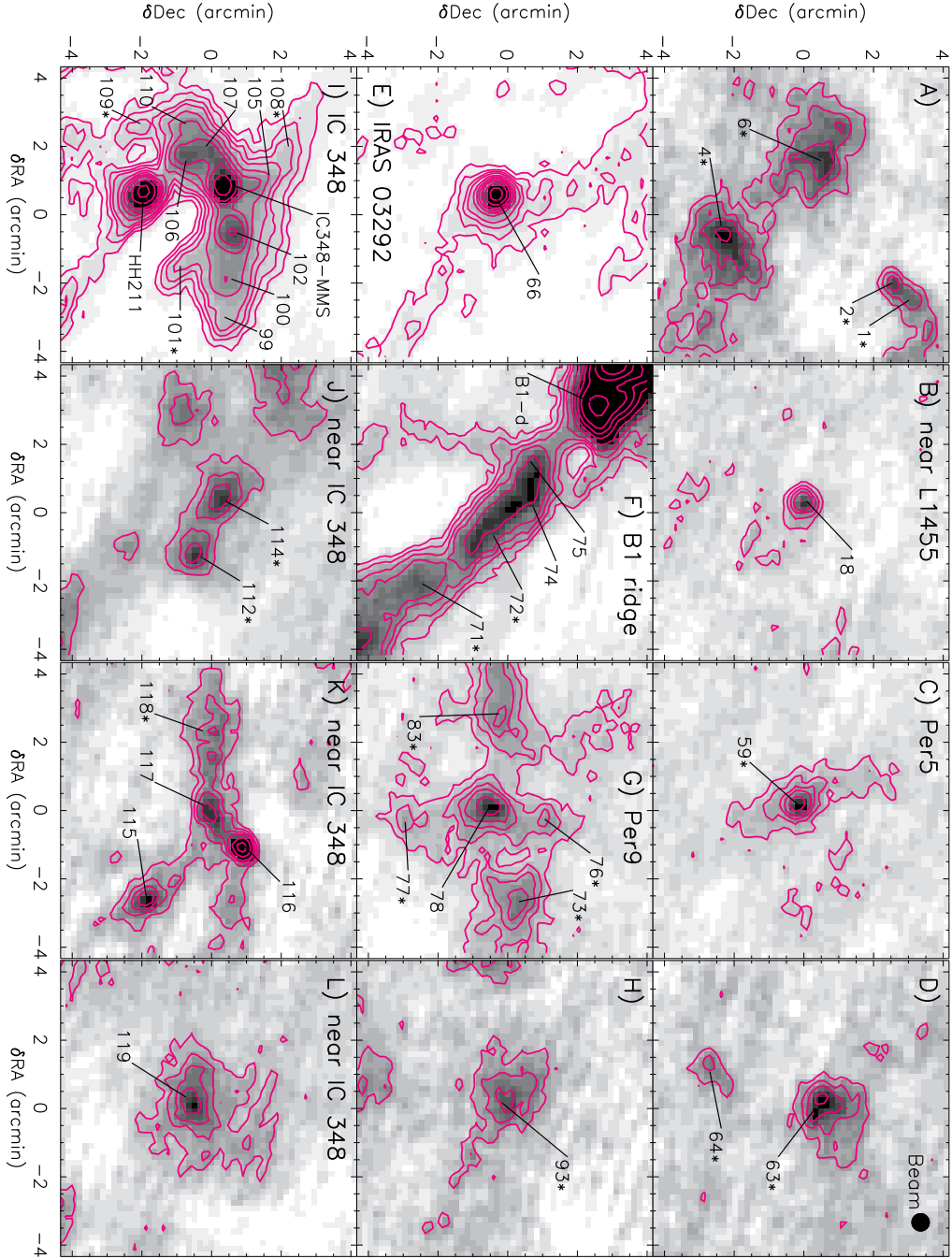}
\caption{Examples of new millimeter detections.  Each image is $8.5\arcmin \times 8.5\arcmin$, and has been smoothed to an effective resolution of $35\arcsec$.  Contours are $(2,4,6,8,12,16,20,25,35,50,75)\sigma$ on the grayscale 1.1~mm map.  Any well know sources are labeled with their common names.  Numbers give the ``Bolo\#'' identification from this paper (Table~\ref{srctab}); those sources with a ``*'' next to the ID were either not covered by or not detected in the $850~\micron$ SCUBA survey of \citet{hatch05}.  Sources range from compact (B) to extended (L) and crowded (I) to isolated (C).  \label{smallsrcsfig}}
\end{figure}

A total of 122 confirmed sources are identified in the optimally filtered map, the locations of which are indicated by small circles on the unfiltered map in Figure~\ref{sourcefig}.  This figure also includes insets of the densest source regions.
Because the RMS varies very little across the map, the lack of sources over large regions of the image is real.  
Many of the sources seen in the 1.1~mm Bolocam map were very recently identified at $850~\micron$ in the SCUBA survey by \citet{hatch05}.  The number of new sources not previously identified at submillimeter or millimeter continuum wavelengths is $\sim 60/122$, most of which are within the region covered by \citet{hatch05} but not detected in that survey due to their somewhat higher mass detection limit ($0.4~M_{\sun}$ (12K) vs. $0.2~M_{\sun}$ (10~K) for the present survey).  A number of these new sources lie within dense NH$_3$ cores \citep[Per3-Per9;][]{ladd94}, or coincide with IRAS sources.  Most are faint, and in the vicinity of known groups, although there are a few more isolated sources.  A selection of new sources chosen to demonstrate the wide range in source properties is shown in Figure~\ref{smallsrcsfig}.  

We additionally identify cores using the 2D Clumpfind routine of \citet{will94}.
Clumpfind is useful for separating sources in crowded regions, where it may be more effective than aperture photometry in defining total flux densities.
It assigns pixels to each source by first contouring the map with a small contour interval ($2\sigma$).  For each peak, contours are followed using a ``friends-of-friends'' algorithm down to one contour below the $5\sigma$ detection limit, or until they run into another source at the same contour level.  Clumpfind makes no assumptions about the clump profile, so the sources need not be gaussian.  

Clumpfind identifies 119 cores in the unfiltered map.   Hereafter we refer to the sample of cores detected using peak finding in the optimally filtered map as ``Peak-find sources'', and the sample identified with Clumpfind as ``Clumpfind sources''.  
The total number of Clumpfind and Peak-find sources is very similar, but there is not a one-to-one correlation between the two sets.  
Fewer faint sources are found with Clumpfind because sources are detected above $5\sigma$ using the RMS \textit{per pixel} of the unfiltered map, which is $\sim \sqrt{3}$ times the RMS \textit{per beam} used to detect sources in the optimally filtered map.  
On the other hand, Clumpfind breaks up many of the brighter Peak-find sources into multiple sources. 
Faint extended sources that one would consider single if examining by eye are often partitioned into multiple sources by Clumpfind, and for this reason we favor the Peak-find source identification and photometry.  We keep Clumpfind calculations for comparison to previous work, as well as to illustrate differences between the two methods.

\subsection{Comparison to Molecular \& $A_V$ Maps}\label{avsect}

\begin{figure}
\plotone{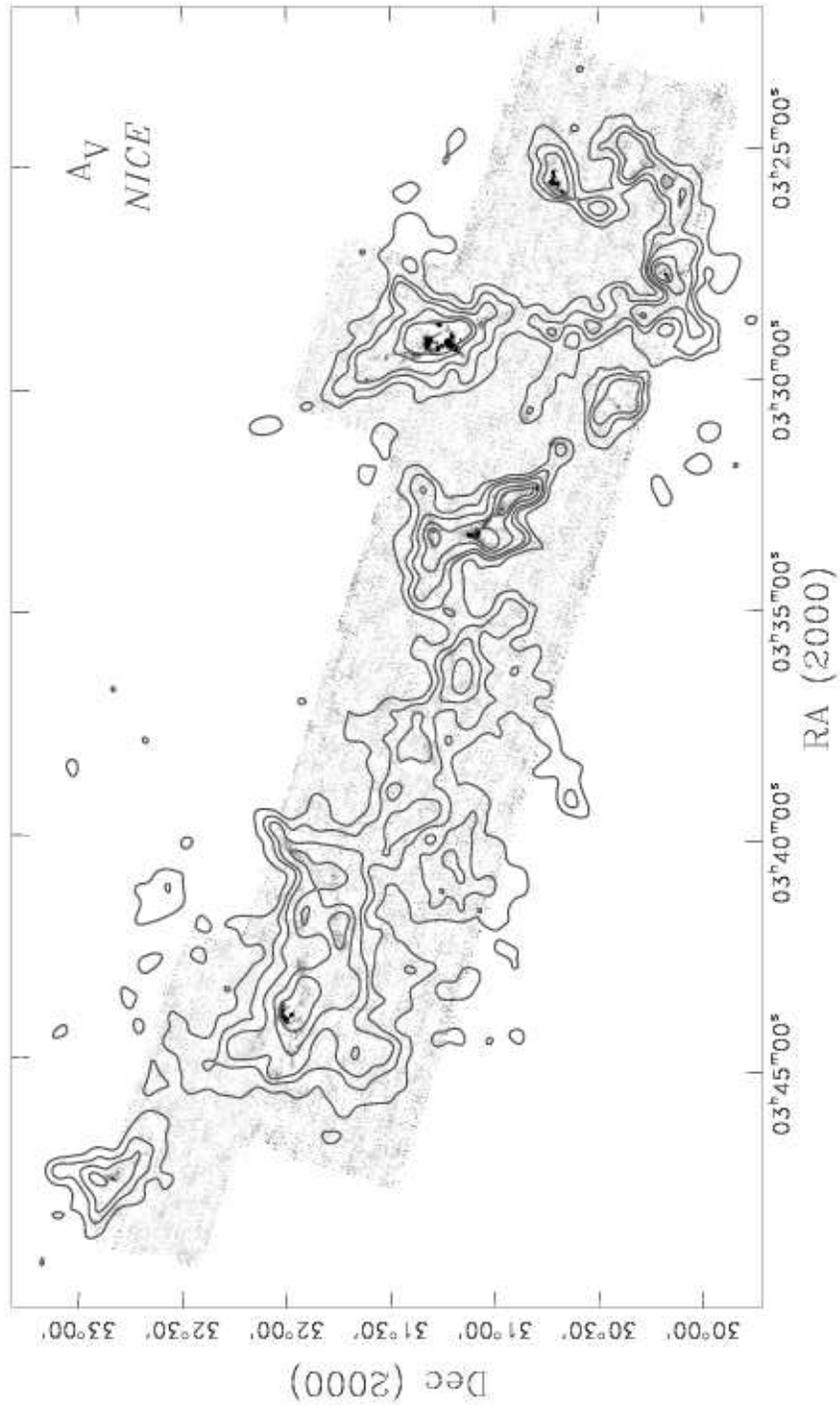}
\caption{Visual extinction ($A_V$) contours calculated from 2MASS data using the ${\it{NICE}}$ method
(see Section~\ref{avsect}), overlaid on the grayscale 1.1~mm map.  Contours in the following plots are $A_V=2,3,4,6,8$~mag with an effective resolution of $5\arcmin$ unless otherwise noted.   
Most 1.1~mm sources lie within relatively high extinction ($A_V\gtrsim5$) peaks.  \label{extnfig}}
\end{figure}

\begin{figure}
\plotone{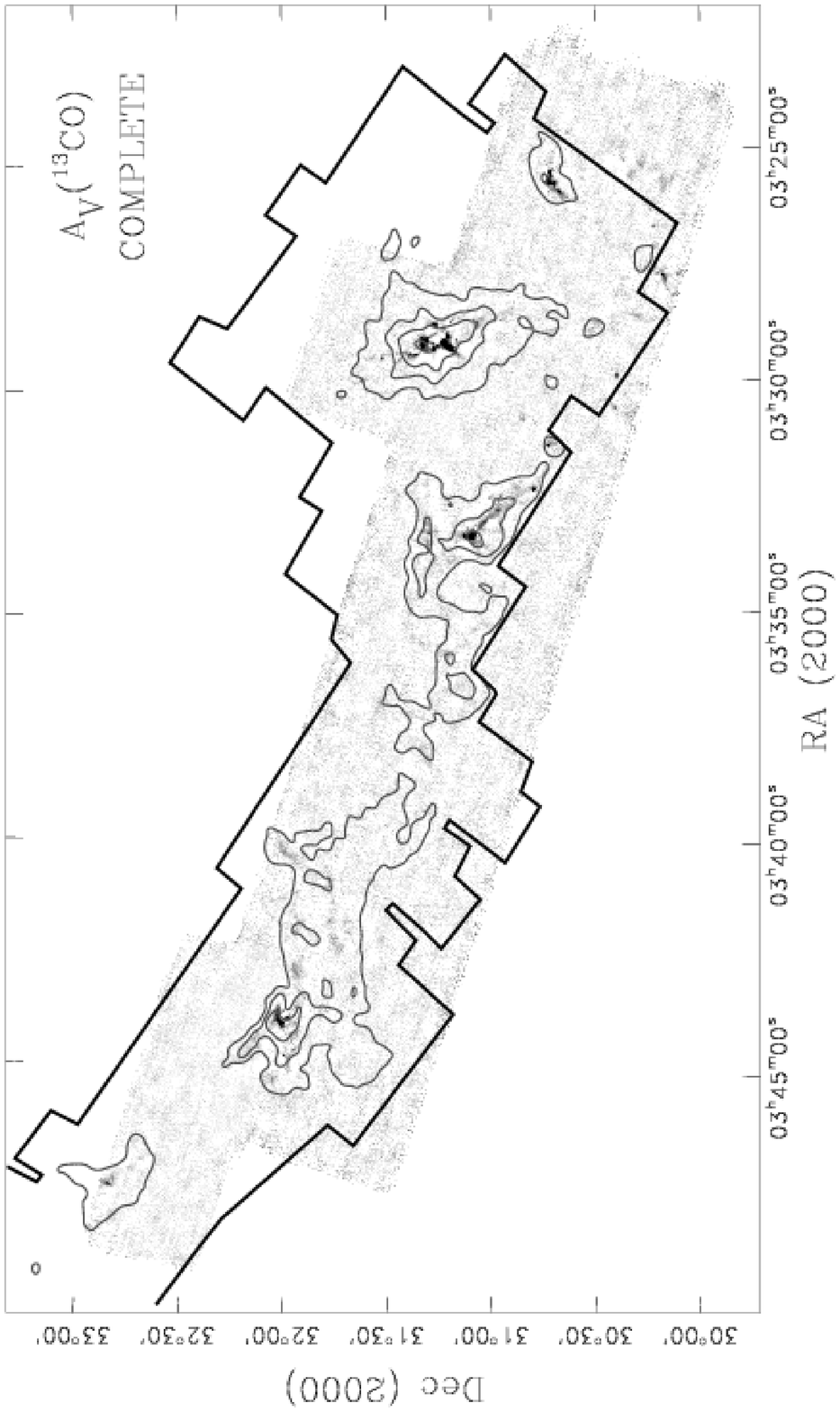}
\caption{$A_V$ contours calculated from the COMPLETE $^{13}$CO map, smoothed to $5\arcmin$ to match the resolution of the {\it{NICE}} extinction map, overlaid on the Bolocam 1.1~mm map.  Thick black lines indicate the observational boundaries of the COMPLETE map.  As a tracer of column density, $^{13}$CO is only accurate where it is optically thin. $A_V$($^{13}$CO) becomes more compact compared to extinction $A_V$, especially around 1.1~mm sources.  \label{av13cofig}}
\end{figure}

\begin{figure}
\plotone{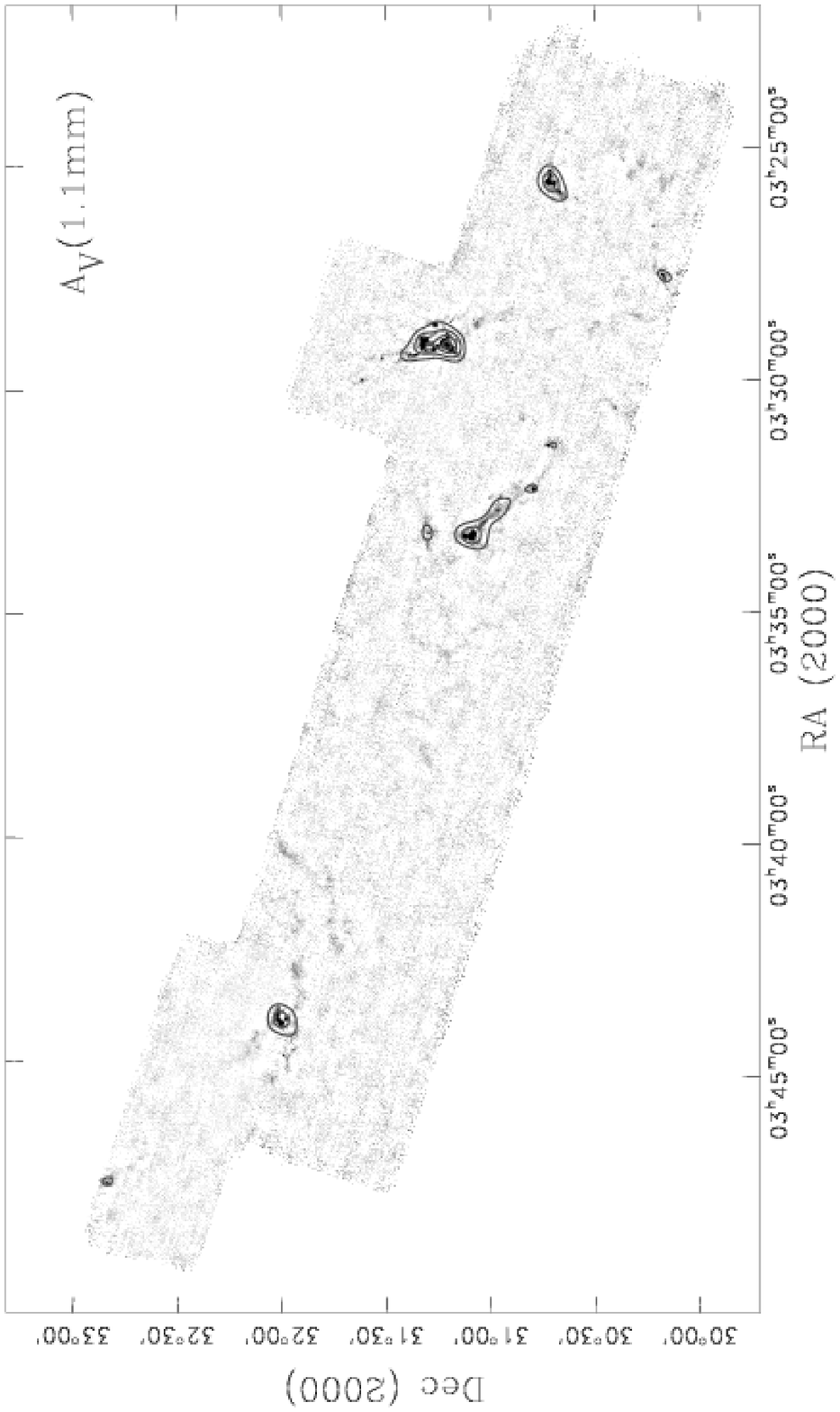}
\caption{$A_V$ contours calculated from the 1.1~mm map, smoothed to $5\arcmin$ resolution.  Contours are $A_V=2,5,10,15,20$~mag.  The 1.1~mm contours are much more compact and extend to higher column density than the other column density tracers, likely because 1.1~mm emission traces only the densest regions of the cloud.  \label{av1mmfig}}
\end{figure}

Large-scale CO and extinction maps of Perseus are available for public use as part of ``The COMPLETE Survey of Star-Forming Regions''\footnote{http://cfa-www.harvard.edu/COMPLETE/} \citep[COMPLETE;][]{good04}.  These observations have been coordinated to cover the Spitzer c2d area of Perseus, therefore they also overlap with the Bolocam observations presented here.
For the following we make use of the COMPLETE $^{13}$CO map, and calculate our own NIR extinction ($A_V$) map.
It is most instructive to compare the 1.1~mm, $A_V$, and $^{13}$CO maps if they are all converted to the same resolution and column density scale.  The three maps are shown as $A_V$ contours overlaid on the grayscale 1.1~mm emission in Figures~\ref{extnfig}-\ref{av1mmfig}.
Contours in all plots are $A_V=2,3,4,6,8$~mag except for the 1.1~mm data (Figure~\ref{av1mmfig}), where $A_V=2,5,10,15,20$~mag to avoid confusion, and smoothed to an effective resolution of $5\arcmin$.

The extinction contours in Figure~\ref{extnfig} were calculated from the
H-K$_s$ color excesses of 2MASS sources using the {\it{NICE}} method
\citep[e.g.][]{lada99,huard06} and convolving with a
Gaussian beam with FWHM of 5$\arcmin$.  This method depends on background stars to
probe the column density through the cloud.  In constructing the extinction
map, we eliminate from the 2MASS catalog most foreground and embedded sources 
that would yield unreliable extinction estimates.  
Foreground sources were identified by color excesses representing a small
line-of-sight visual extinction, $A_{V,LOS} \le\ 3$ mag, within a
5$\arcmin\ \times\ $5$\arcmin$ region exhibiting a large mean extinction
of $A_V \ge $ 8 mag.  Embedded sources were identified as those
sources positioned at least 1$\sigma$ redward of the reddened
main-sequence region on a (J-H, H-K) color-color diagram.  We
use only those sources with reasonably good photometry at J, H, and K$_s$
(photometric errors less than 0.5, 0.2, and 0.2 mag, respectively)
to construct the extinction map.  The color-excesses were computed
assuming that the intrinsic color of non-extincted stars in off-cloud
fields are representative of the background star population.  
These two off-cloud regions, having a total of more than 5700 stars, were $1\fdg 5 \times 0\fdg4$ fields centered on $\alpha  = 03^h56^m48.0^s,~\delta  = 35^\circ 06^\prime00^{\prime\prime}$ and $\alpha  = 03^h49^m12.0^s,~ \delta  = 27^\circ 54^\prime00^{\prime\prime}$ (J2000.0).

The NIR-derived extinction contours in Figure~8 verify that the Bolocam
1.1~mm map covers most of the $A_V\ge2$~mag area.  All bright
1.1~mm sources occur in regions of high extinction, but the reverse is not 
necessarily true; high extinction does not guarantee the presence either of 
young protostars or dense millimeter cores with strong 1.1~mm emission.  Note 
that the extinction map saturates around $A_V\gtrsim10$~mag.

Where $^{13}$CO is optically thin it provides another measure of column density.  This may only be true for a small range of densities, however, as $^{13}$CO can be under-excited at small densities and becomes optically thick at $A_V\sim6$~mag for a typical linewidth of $\Delta v \sim 1$~km~s$^{-1}$. 
To convert $^{13}$CO integrated intensity to H$_2$ column density, we assume LTE, $\tau \ll 1$, and an excitation temperature $T_x=10$~K, following \citet{dick78}. $A_V$ contours are calculated taking $N_{LTE}(^{13}$CO$) = 2.5 \times 10^{15} (A_V-0.8)$~cm$^{-2}$ in Perseus for $1<A_V<5$~mag \citep{BC86}, and shown in Figure~\ref{av13cofig}.  
The $A_V$($^{13}$CO) and {\it{NICE}} extinction maps display the same general morphology.  Although the $^{13}$CO contours do not reach peak densities as high as those in the {\it{NICE}} $A_V$ map, they do become more compact compared to the NIR extinction derived $A_V$, especially around 1.1~mm sources.

H$_2$ column density is calculated from the 1.1~mm map using a conversion from 1.1~mm flux density per beam ($S_{1.1mm}^{beam}$) of
\begin{equation}
N(\mathrm{H}_2) = \frac{S_{\nu}^{beam}}{\Omega_{beam} \mu_{H_2} m_H \kappa_{\nu} B_{\nu}(T_D)}. \label{aveq}
\end{equation}
Here $\Omega_{beam}$ is the beam solid angle, $m_H$ is the mass of hydrogen, $\kappa_{1.1mm} = 0.0114$~cm$^2$g$^{-1}$ is the dust opacity per gram of gas, $B_{\nu}$ is the Planck function, $T_D$=10~K is the dust temperature, and $\mu_{H_2}=2.8$ is the mean molecular weight per H$_2$ molecule, which is the relevant quality for conversion to extinction.  
We extrapolate $\kappa_{1.1mm}$ from \citet[][Table~1 column 5, hereafter OH5]{oh94} for dust grains with thin ice mantles, coagulated for $10^5$ years at a gas density of $10^6$~cm$^{-3}$.  A gas to dust ratio of 100 is assumed.  OH5 dust has found to be the best choice for star-forming cores by several authors using radiative transfer modeling \citep{evans01, shir02, young03}.  
Column densities calculated from thermal emission depend on $T_D$, which varies with position.  Thus there may be local discrepancies with other tracers when $T_D$ is not independently known, but agreement should be good overall.

We assume a conversion from column density to $A_V$ of $N($H$_2)/A_V = 0.94 \times 10^{21}$ cm$^{2}$~mag$^{-1}$ \citep{flw82}.  The resulting $A_V$(1.1~mm) contours in Figure~\ref{av1mmfig} clearly demonstrate that the 1.1~mm emission is considerably more compact compared to the other column density tracers.  It appears that only the densest material is traced by the 1.1~mm emission, as manifested in both the compact nature of the $A_{V}$(1.1~mm) contours and the high $A_V$ reached in the bright cores ($A_{V,max}\sim17$ at $5\arcmin$ resolution).  The $5\sigma$ detection limit per $31\arcsec$ beam corresponds to a minimum $A_V\sim 5$~mag, so it is not surprising that there is not 1.1~mm emission seen in the lower $A_V$ regions of the $A_V$ and $^{13}$CO maps.
We note, however, that our simulations indicate that the Bolocam map may not be sensitive to more extended emission ($\gtrsim 4\arcmin$) even if it is present in the cloud.

\subsection{Source statistics}\label{statsect}

Source identifications, positions, peak flux densities ($S_{peak}$), and signal to noise (S/N) for the 122 1.1~mm sources identified in the Bolocam map are listed in Table~\ref{srctab}.  
The S/N is measured in the optimally filtered map, because this is the S/N that determines detection.
Photometry and all other source properties are measured in the unfiltered, surface brightness-normalized map. 
$S_{peak}$ is the peak pixel value in mJy/beam as measured in the $10\arcsec$/pixel map.  
The uncertainty in $S_{peak}$ is the local RMS/beam, calculated as described in Section~\ref{idsect}.  
There is an additional systematic error of $\sim15$\% associated with all flux densities, from the absolute calibration uncertainty (10\%) and the systematic bias remaining after iterative mapping ($\sim5$\%, see Section~\ref{itmapsect}).
Table~\ref{srctab} also lists the most commonly used name from the literature for known sources, and indicates if the 1.1~mm source is coincident (within $40\arcsec$) with a MIPS $24~\micron$ source from the c2d database (Rebull et al. 2005, in preparation).  The c2d MIPS data for some regions of Perseus are not yet available, in which case ``N/A'' is listed in the table.  For these sources, the presence of an IRAS source (within $1\arcmin$) is indicated if appropriate.  

Table~\ref{phottab} lists photometry in fixed apertures  of diameter $40\arcsec$, $80\arcsec$, and $120\arcsec$, the total integrated flux density ($S_{\nu}$), total mass, peak $A_V$, full-width at half maximum (FWHM) sizes along the major and minor axes, position angle (PA, measured east of north), mean particle density $\mean{n}$, and morphology descriptions for each source.   Integrated flux densities are measured assuming a sky value of zero, and corrected for the gaussian beam so that a point source has the same integrated flux density in all apertures with diameter greater than the beam.  
No integrated flux density is given if the distance to the nearest neighboring source is smaller than the aperture diameter.  

To calculate the total flux density, we compute integrated flux densities in apertures of $30\arcsec-160\arcsec$ in intervals of $10\arcsec$, and use the largest aperture diameter that is smaller than the distance to the nearest neighbor. An aperture of $120\arcsec$ is generally sufficient unless the source FWHM is $>100\arcsec$, in which case apertures up to $160\arcsec$ are used.  
If the flux density decreases in larger apertures (due to residual negative artifacts), then the aperture giving the largest flux density is used.   
Uncertainties in all integrated flux densities are $\sigma_{ap} = \sigma_{beam} \sqrt{ A_{ap}/A_{beam}}$, where $\sigma_{beam}$ is the local RMS/beam and ($A_{ap}$, $A_{beam}$) are the aperture and beam areas respectively, and do not include an additional 15\% systematic uncertainty. 
The FWHM and PA are from the best fit elliptical gaussian after masking out nearby sources using a mask radius equal to half the distance to the nearest neighbor.  The errors given are the formal fitting errors and do not include uncertainties due to noise or residual cleaning effects, which are of order $10-15\%$ for the FWHM and $\sim 5\degr$ for the PA.

\begin{figure}
%\plotone{f11.eps}
\plotone{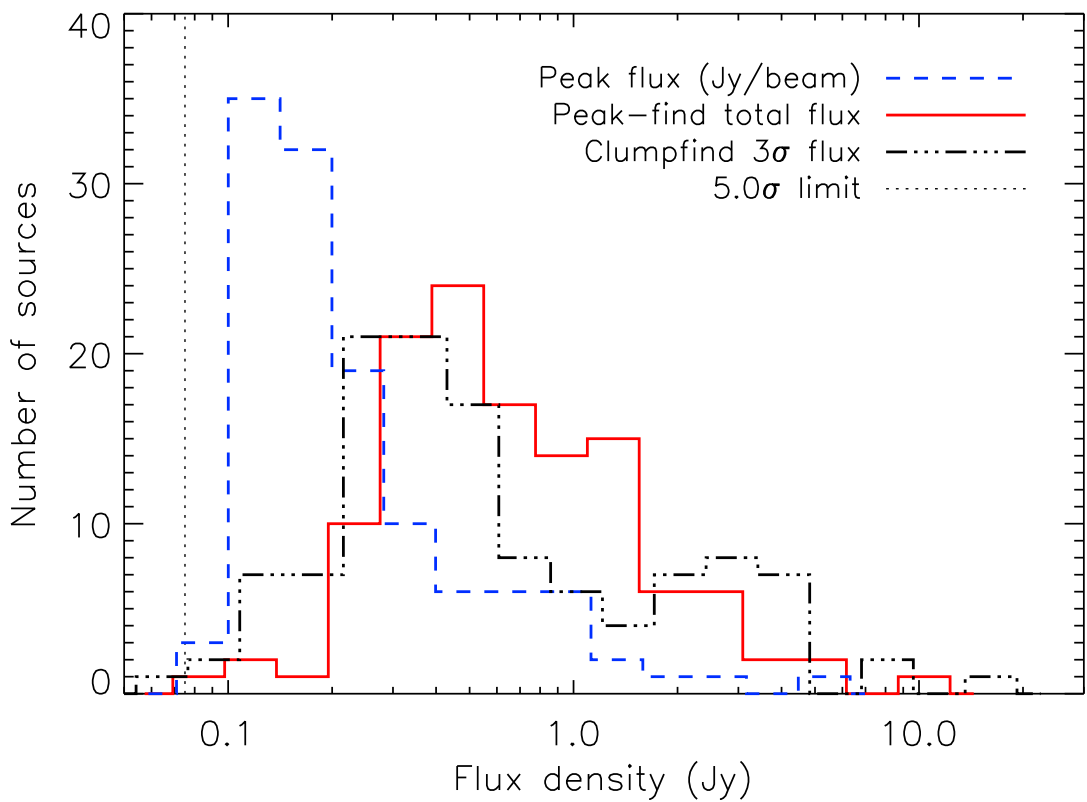}
\caption{Distribution of source peak flux densities (dashed line, mJy/beam) and total flux densities (solid line, Jy) from aperture photometry.  Clumpfind $3\sigma$ flux densities are also shown for comparison (dash-dot line, Jy).  The vertical dotted line is the $5 \sigma$ peak detection limit.  The peak flux density distribution has a mean of $0.35$~Jy/beam and dispersion of $0.56$~Jy/beam ($0.35\pm0.56$), and the total flux density distribution a mean of $0.96\pm1.21$~Jy.  The mean flux from Clumpfind ($1.13\pm1.80$~Jy) is somewhat larger than that from aperture photometry.  \label{fluxfig}}
\end{figure}

The distribution of source flux densities is shown in Figure~\ref{fluxfig}.  In addition to peak flux densities (Jy/beam) and total flux densities (Jy) from aperture photometry of Peak-find sources, Clumpfind $3\sigma$ flux densities (Jy) are also shown.
Clumpfind  flux densities are calculated from the source map generated by Clumpfind as the sum over all $>3\sigma$ pixels assigned to a given source, divided by the beam area.
We find empirically that Clumpfind is good at separating, and calculating the correct flux density for, bright, crowded sources.  It is not, however, very effective at determining the total flux density of isolated faint or extended sources.   For example, faint sources will have a smaller total flux density than for aperture photometry because Clumpfind only includes pixels above $3\sigma$.

The large average size of sources accounts for the fact that the mean total flux density ($0.96\pm1.21$ Jy; hereafter, numbers quoted are mean $\pm$ dispersion in the sample, \textit{not} mean $\pm$ error in the mean) is significantly larger than the mean peak flux density ($0.35\pm0.56$ Jy/beam).  
Clumpfind sources tend to have slightly larger $3\sigma$ flux densities ($1.13\pm1.80$~Jy), in part because bright, crowded sources may be integrated over a larger area than with aperture photometry, and in part because one Clumpfind source sometimes encompasses more than one Peak-find source.

\begin{figure}
%\plotone{f12.eps}
\plotone{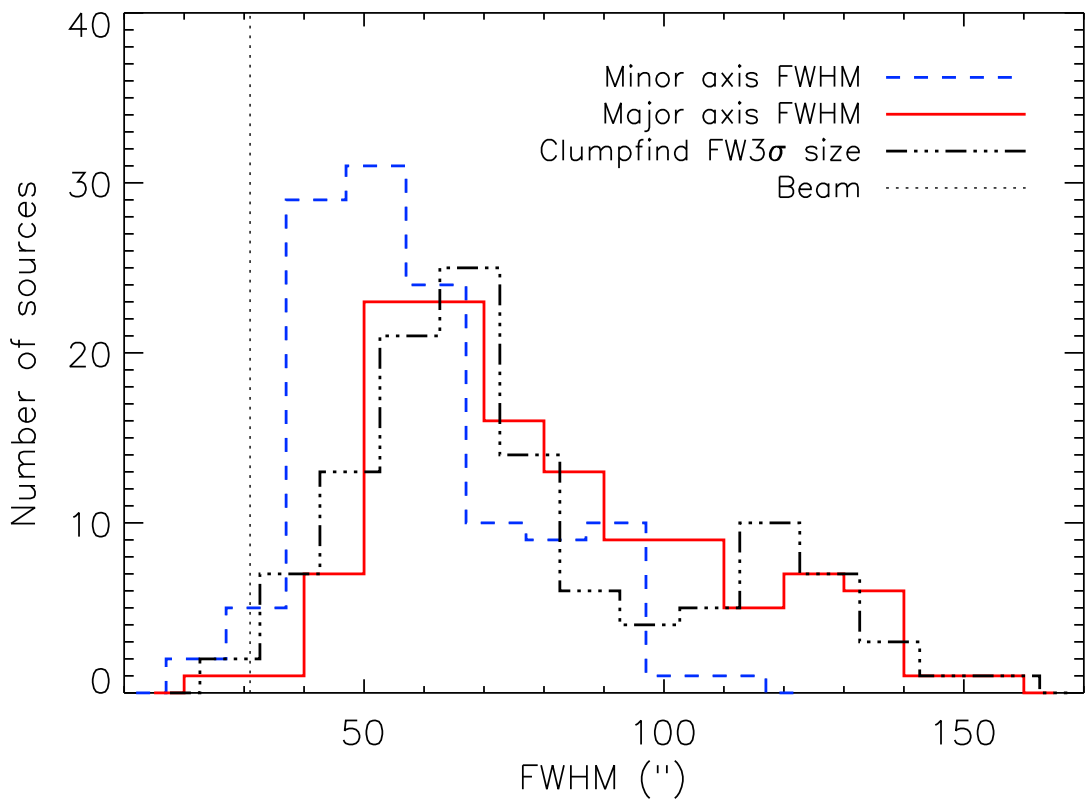}
\caption{Distribution of minor axis (dashed line) and major axis (solid line) FWHM sizes, as determined from an elliptical gaussian fit.  The beam size is indicated by the dotted line.  The mean minor axis FWHM is $58\arcsec \pm 17\arcsec$, and the mean major axis FWHM is $80\arcsec \pm 27\arcsec$.  
Full-width at $3\sigma$ sizes for Clumpfind sources are also shown (dash-dot line), where (FW$3\sigma=2\sqrt{N_{pix}A_{pix}/\pi}$.  The mean size from Clumpfind ($77\arcsec\pm30\arcsec$) is larger than the mean FWHM ($68\arcsec\pm20\arcsec$), as expected because the Clumpfind size is measured at the $3 \sigma$ contour rather than at the half-max.  \label{sizefig}}
\end{figure}

\begin{figure}
\plotone{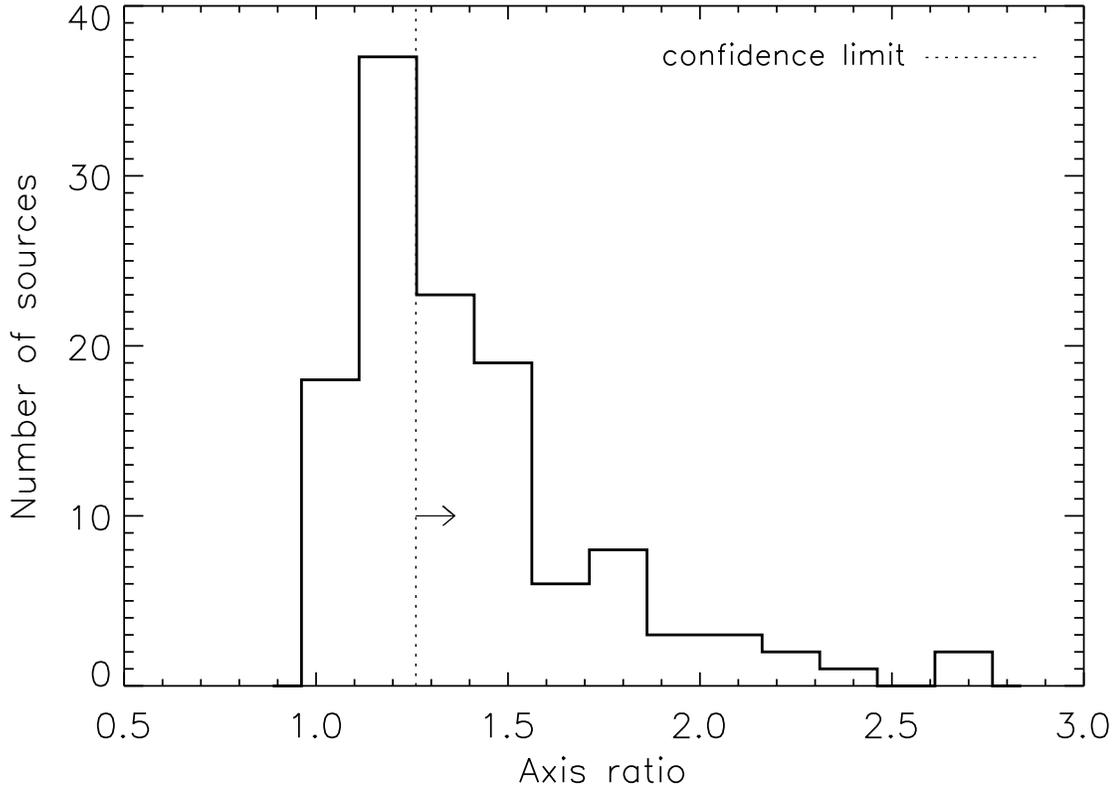}
\caption{Distribution of source axis ratios.  The mean axis ratio is 1.4, and the sample contains some very elongated sources with axis ratios $\ge 2$.  We find that measured axis ratios $<1.2$ are unreliable based on Monte Carlo tests. \label{axisfig}}
\end{figure}

Figure~\ref{sizefig} gives the distribution of major and minor FWHM sizes, as well as the full-width at $3\sigma$ (FW$3\sigma$) sizes of Clumpfind sources, defined as FW$3\sigma=2\sqrt{N_{pix}A_{pix}/\pi}$, where $N_{pix}$ is the total number of pixels assigned to a source and $A_{pix}$ is the pixel area.
The average minor axis FWHM is $58\arcsec \pm 17\arcsec$, the average major axis FWHM is $80\arcsec \pm 27\arcsec$, and the average mean FWHM is $68\arcsec \pm 20\arcsec$.  Clumpfind sources have a large average FW$3\sigma$ of $77\arcsec \pm 30\arcsec$, as expected given that the size is measured at $3\sigma$ rather than at half-max.  There are very few sources near the resolution limit of $31\arcsec$.  Most sources are somewhat elongated as well as extended; the average axis ratio is 1.4, and the distribution (Figure~\ref{axisfig}) extends to large axis ratios $>2$.  Note, however, that measured axis ratios $<1.2$ are found to be unreliable based on simulations, and should be considered indistinguishable from an axis ratio of unity.

The total mass $M$ of gas and dust in a core is proportional to the total flux density $S_{\nu}$, assuming the dust emission at 1.1~mm is optically thin and both the dust temperature and opacity are independent of position within a core:
\begin{equation}
M = \frac{d^2 S_{\nu}}{B_{\nu}(T_D) \kappa_{\nu}}, \label{masseq}
\end{equation}
where $\kappa_{1.1mm} = 0.0114$~cm$^2$~g$^{-1}$ is the dust opacity, $d=250$~pc is the distance, and $T_D$ is the dust temperature. 
Although the millimeter emission arises only from the dust, we can infer the total mass of gas and dust by assuming a gas to dust mass ratio of 100, which is included in $\kappa_{1.1mm}$.   For the masses in Table~\ref{phottab}, we assume a single dust temperature $T_D=10$~K for all sources.  The uncertainties given are from the uncertainty in the total flux density only.  Other sources of error from $\kappa$, $T_D$, and $d$ (up to a factor of 4 or more) are discussed in Section~\ref{masssect}.

For dense regions without internal heating, the mean temperature is about 10~K, warmer on the outside and colder on the inside \citep{evans01}. Centrally heated cores will be warmer on the inside, but much of the mass is located at low temperatures. \citet{shir02} and \citep{young03} found good agreement with masses of Class~0 and Class~I sources determined from detailed models using $T_D \sim 15$~K. Taking 10~K is a reasonable compromise to cover both prestellar and protostellar sources, but keep in mind that it will overestimate the masses of the latter by a factor of $2-3$.

The peak $A_V$ in Table~\ref{phottab} is calculated from the peak 1.1~mm flux density $S_{peak}$ as in Equation~\ref{aveq}.  The average peak $A_V$ of the sample is 24.7~mag.
The mean particle density for each source is estimated as $\mean{n} = M/(4/3 \pi R^3 \mu)$, where $M$ is the total mass, $R$ is the mean deconvolved HWHM size, and $\mu_p=2.33$ is the mean molecular weight per particle.  The average $\mean{n}$ of the sample is $4.3\times 10^5$~cm$^{-3}$.
The morphology keywords listed indicate if the source is multiple (within $3\arcmin$ of another source), extended (major axis FW at $2\sigma > 1\arcmin$), elongated (axis ratio at $4\sigma > 1.2$), round (axis ratio at $4\sigma < 1.2$), or weak (peak flux densities less than 5 times the RMS per pixel in the unfiltered map). 
Monte Carlo simulations (Section~\ref{itmapsect}) indicate that we cannot recover structures larger than $\sim 200\arcsec$, and we do not find any sources larger than $\sim 120\arcsec$ in the real map.  We do not resolve any source pairs closer than $32\arcsec$, close to the minimum separation of $30\arcsec$ required by the peak finding algorithm.

\section{Discussion}\label{discsect}

\subsection{Completeness \& the mass vs. size distribution}

\begin{figure}
\plotone{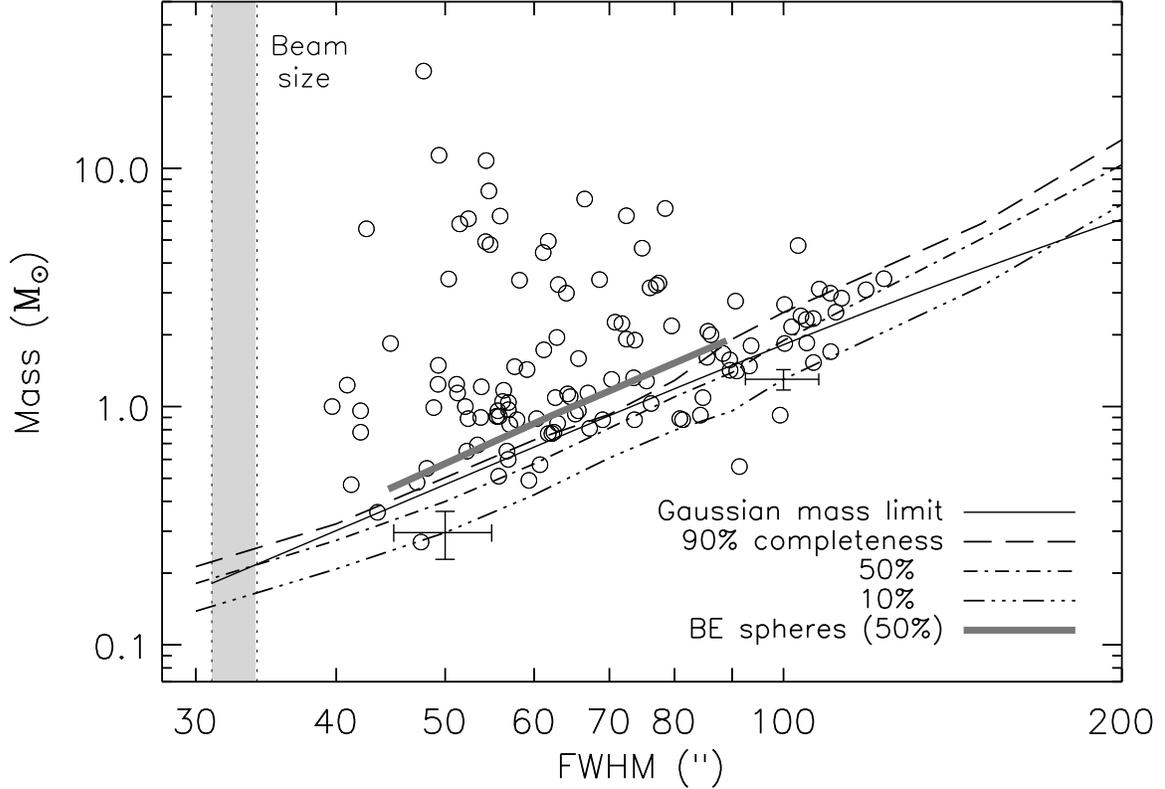}
\caption{Total mass vs. FWHM size for Peak-find sources. 
The solid line denotes the analytic mass detection limit as a function of size ($M\propto R^2$) for gaussian sources.  Empirical 10\%, 50\%, and 90\% completeness curves are also shown, derived using Monte Carlos with simulated sources and taking into account the effects of cleaning, iterative mapping, and optimal filtering.  50\% completeness for Bonnor-Ebert spheres (thick shaded line) were similarly determined using BE models with $n_c = (8\times10^4, 8\times10^4, 9.5\times10^4)$~cm$^{-3}$ and $r_o = (8\times 10^3$, $1.5\times 10^4$, $3\times 10^4)$~AU.  Representative error bars for $50\arcsec$ and $100\arcsec$ FWHM sources near the detection limit are shown, as estimated from the results of Monte Carlo simulations.  Note the lack of sources near the resolution limit, which cannot be entirely be accounted for by pointing errors of $\lesssim7\arcsec$, although the pointing-smeared beam could be as large as $34\arcsec$ (shaded region). \label{mvsfig1}}
\end{figure}

The distribution of total mass vs. FWHM size is shown in Figure~\ref{mvsfig1}.
The solid line shows the expected mass detection limit for gaussian sources assuming a simple scaling with source area.  The gaussian mass limit varies with source size ($M_{lim} \propto \mathrm{size}^2$) because our ability to detect sources is based on the peak flux density above the noise ($5\sigma$), whereas the mass is calculated from the total flux density ($S_{\nu} \propto \mathrm{size}^2$).  
The real mass completeness limit is more complicated, even for gaussian sources, due to our reduction and detection techniques.  We also show, therefore, empirical mass detection limits as a function of size for 10\%, 50\% and 90\% completeness.

Completeness is determined by Monte-Carlo simulations, taking into account effects from cleaning, iterative mapping, and optimal filtering.   Simulated sources of varying total mass and size are inserted into an empty region of the real map before cleaning and iterative mapping.  The 10\% completeness limit is the mass at which 10\% of the simulated sources are detected in the optimally filtered map.
Nearly all detected sources lie above the 10\% completeness limit, as expected.  Note that the completeness curve represent the true, not measured, mass and size of input simulated sources.  

Typical measurement error bars in $M$ and FWHM are shown for $50\arcsec$ and $100\arcsec$ FWHM sources near the detection limit.  For these, $\sigma_M$ are from the uncertainty in the integrated flux, and $\sigma_{\mathrm{FWHM}}$ are estimated from simulations.  The maximum size of the pointing-smeared beam is indicated by a shaded band. 
Using the optimally filtered map to detect sources tends to lower the mass limit for sources with FWHM~$>30\arcsec$ because the peak is enhanced by the filter.  Conversely, very large sources ($>100\arcsec$) tend to have a higher mass limit because they are not fully recovered by iterative mapping.  Both effects are illustrated by the empirical 50\% completeness curve, which falls below the gaussian 50\% completeness line for FWHM~$\sim 40\arcsec-60\arcsec$, and above for FWHM~$\ga100\arcsec$.  

There is an additional complication due to the fact that the sources may not be gaussian in shape.  Without knowing the true source structure this effect is difficult to quantify.  We explore the effects of different source structures by running completeness tests for sources with a Bonnor-Ebert (BE) sphere profile \citep{ebert55,bonn56}, which has been found by several authors to be a good representation of the structure of some prestellar cores \citep{sg05, kirk05, evans01,wt94}. 
The 50\% completeness limits for BE spheres are indicated in Figure~\ref{mvsfig1} by a thick shaded line.  
FWHM sizes of BE spheres, measured with a gaussian fit for consistency with other limits, are set by the chosen outer radius ($r_o$) in the BE model.  The total mass is determined by a combination of $r_o$ and the central density $n_c$, and detection primarily by $n_c$.
The BE sphere models used to calculate completeness have $n_c = (8\times10^4,8\times10^4,9.5\times10^4)$~cm$^{-3}$, $r_o = (8\times 10^3,1.5\times10^4,3\times10^4)$~AU, and $M_{tot} = (0.45, 0.91, 1.89)~ M_{\sun}$.

\begin{figure}
\plotone{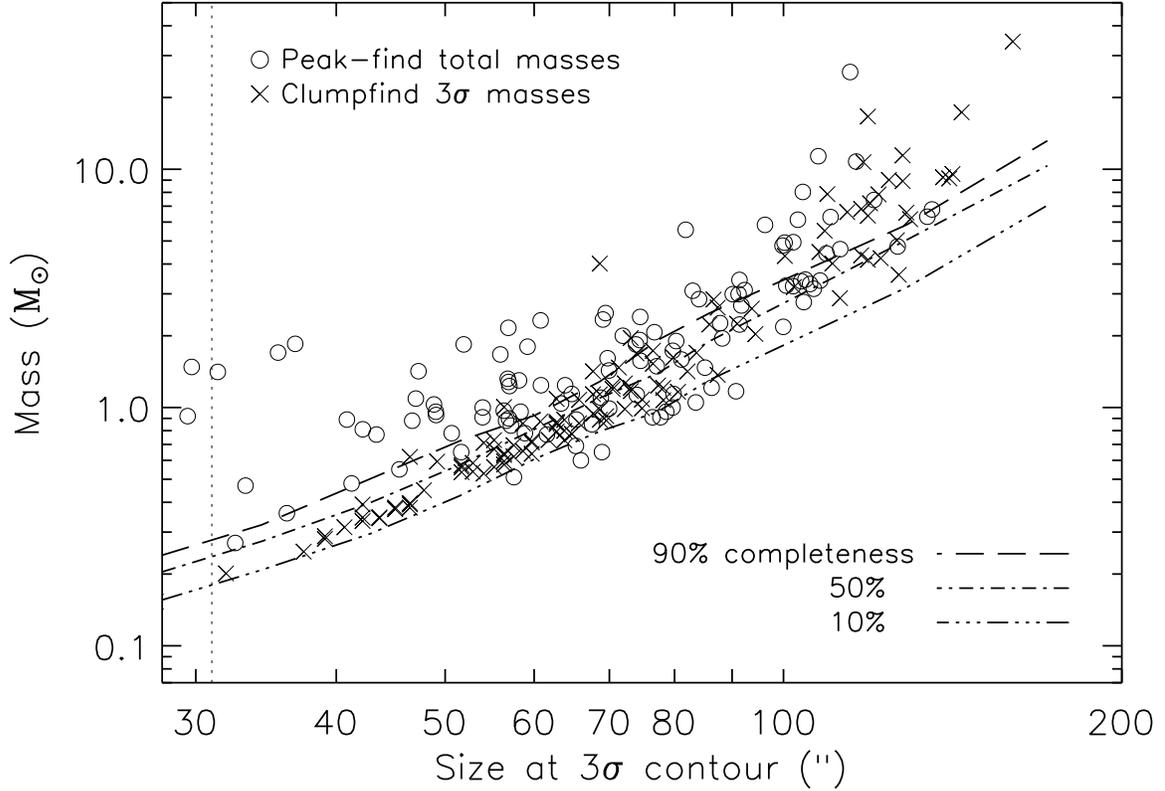}
\caption{Total mass vs. FW$3\sigma$ size for Peak-find and Clumpfind sources. 
The difference in the two distributions lies in the total flux calculation: aperture photometry for Peak-find, and integrating over the $>3\sigma$ pixels for Clumpfind.  
Sizes for Peak-find sources are computed by scaling the FWHM to the $3\sigma$ contour (FW$3\sigma = $FWHM$\times \sqrt{\ln{(S_{\nu,peak}/3\sigma})/\ln{2}}$), assuming the source is gaussian.  The Clumpfind size is based on the total pixel area above the $3\sigma$ contour (FW$3\sigma=2\sqrt{N_{pix}A_{pix}/\pi}$).  Completeness curves are as in Figure~\ref{mvsfig1}, scaled to the $3\sigma$ size.  The smaller apparent scatter compared to Figure~\ref{mvsfig1} is not real, but rather a consequence of the different size definitions used. \label{mvsfig2}}
\end{figure}

Figure~\ref{mvsfig2} is similar to Figure~\ref{mvsfig1} except that the mass versus FW$3\sigma$ size is plotted for both Peak-find and Clumpfind sources.  
For Peak-find sources, FW$3\sigma$ is scaled from the FWHM assuming a gaussian profile (FW$3\sigma = $FWHM$\times \sqrt{\ln{(S_{peak}/3\sigma})/\ln{2}}$). 
Masses for the two are calculated slightly differently: the Peak-find mass is the total mass from aperture photometry, whereas the Clumpfind $3\sigma$ mass is from the integrated flux density within the $3\sigma$ contour. 
The total mass for Peak-find sources (circles) has not changed from Figure~\ref{mvsfig1}, but the distribution of mass vs. size looks very different  because the FW$3\sigma$ size tends to be larger than the FWHM for bright sources (by about a factor of 2, because the $3\sigma$ contour is well below the half-max), but smaller than the FWHM for faint sources. 
The apparent decrease in scatter is not real, but simply an effect of the size definition used. 
For faint sources the Clumpfind $3\sigma$ mass is smaller than the Peak-find aperture photometry mass, because Clumpfind only integrates down to $3\sigma$, which corresponds to a small aperture for faint sources near the $5\sigma$ detection limit. 

The appearance of the mass vs. size distribution depends very strongly on the definitions of both size and total mass, as illustrated by Figures~\ref{mvsfig1} and ~\ref{mvsfig2}, making comparisons between plots calculated in different ways quite deceptive.  For example, the $3\sigma$ mass vs. $3\sigma$ size distribution for Clumpfind sources in Figure~\ref{mvsfig2} (crosses) is very similar to that in Ophiuchus from \citet{john00}, but comparing the Ophiuchus plot to the Peak-find total mass vs. FWHM size would make the two distributions seem very different.

We find it significant that there seem to be very few point sources in Perseus at 1.1~mm.  This is most clearly illustrated in Figure~\ref{mvsfig1}, which demonstrates an obvious paucity of both faint and bright sources between FWHM$\sim30\arcsec-50\arcsec$ (although there are low mass, compact sources in Figure~\ref{mvsfig2}, these are artificially created by the $3\sigma$ cutoff).
Given that the pointing-smeared beam is no larger than $34\arcsec$ (for pointing errors $\lesssim 7\arcsec$) and the average FWHM is more than twice the beam size ($68\arcsec$ vs. $31\arcsec$), the majority of sources are significantly extended.  In fact, the average deconvolved size is $61\arcsec = 1.5 \times 10^4$~AU.  This mean size is inconsistent with descriptions of cores as truncated spheres, which have been used to model very low mass cores.  A truncated power law model for an $0.3~M_{\sun}$ core requires an outer radius of $2\times 10^3$~AU \citep{ye05}, which would be a point source at the resolution of Bolocam.    
Our mass limit is $0.18~M_{\sun}$ for a point source at $T_D=10$~K, so we should be sensitive to such compact, low mass cores if they are present in Perseus.  

\subsection{The 1.1~mm Mass Function}\label{masssect}

\begin{figure}
%\plotone{f16.eps}
\plotone{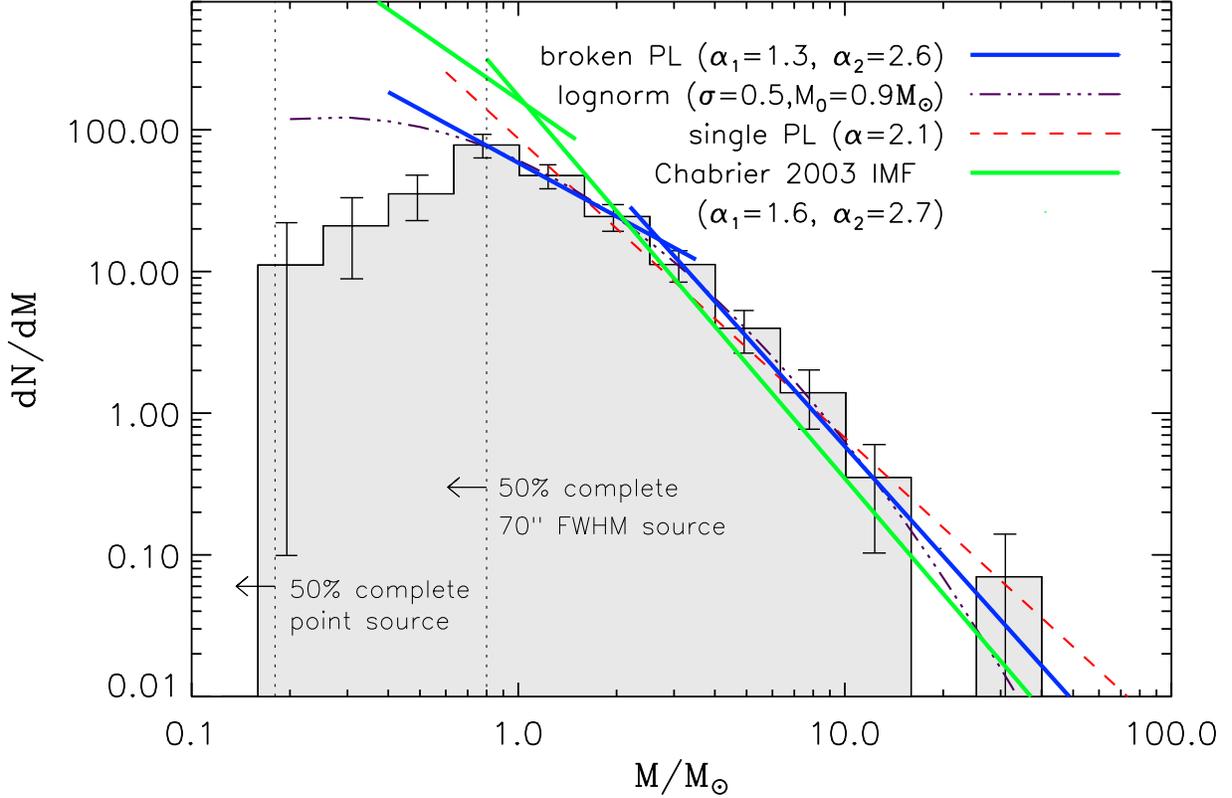}
\caption{Differential mass function $dN/dM$ for masses calculated from aperture photometry and a single dust temperature $T_D=10$~K.  The 50\% completeness limit is $0.18~M_{\sun}$ ($T_D=10$~K) for a point source, and $0.8~M_{\sun}$ for a $70\arcsec$ FWHM source, the average size of the sample.  Assuming a broken power law of the form $N(M) \propto M^{-\alpha}$, the best fit slopes are $\alpha_1=1.3\pm0.3$ ($0.5M_{\sun}<M<2.5M_{\sun}$) and $\alpha_2=2.6\pm0.3$ ($M>2.5~M_{\sun}$). The slope, but not the break mass, is very similar to the local IMF: $\alpha_1=1.6$ ($M<1~M_{\sun}$), $\alpha_2=2.7$ ($M>1~M_{\sun}$) \citep{chab03}.   The data for $M>0.8~M_{\sun}$, where the distribution is not affected by completeness, is also well fitted by a lognormal with $\sigma=0.5\pm0.1$, $M_0=0.9\pm0.4~M_{\sun}$.   The best fit single power law is $\alpha=2.1\pm0.1$ for $M>0.5~M_{\sun}$.  For comparison, $850~\micron$ sources in Ophiuchus were found to have $\alpha_1 \sim 1.5$ below $1~M_{\sun}$ and $\alpha_2 \sim 2.5$ above $1~M_{\sun}$ \citep{john00}.  
\label{difffig}}
\end{figure}

The differential mass function $dN/dM$ for all 122 1.1~mm sources is shown in Figure~\ref{difffig}.   The average mass of the sample is $\mean{M}=2.3M_{\sun}$, with individual masses ranging from $0.2$ to $26M_{\sun}$.  
A dust opacity of $\kappa_{1.1mm} = 0.0114$~cm$^2$~g$^{-1}$ (OH5), temperature of $T_D=10$~K, and  distance of $d=250$~pc are assumed for all masses.  
Error bars are $\sqrt{N}$ statistical errors, and are meant to demonstrate the typical uncertainties from photometry, but do not include errors due to uncertainties in the distance or dust properties.  Completeness becomes important around $0.2~M_{\sun}$ (50\% completeness) for point sources, and around $0.8~M_{\sun}$ for sources with a FWHM of $\sim 70\arcsec$, the average FWHM of the sample.  The mass distrubution is not corrected for incompleteness, therefore the turnover below $\sim 0.8~M_{\sun}$ may not be real. 

The best fit to the mass function is a broken power law  
\begin{equation}
\frac{dN}{dM} \propto M^{-\alpha}
\end{equation}
with $\alpha_1=1.3\pm0.3$ ($0.5M_{\sun}<M<2.5M_{\sun}$) and $\alpha_2=2.6\pm0.3$ ($M>2.5~M_{\sun}$).  
This result has a reduced chi-squared of $\tilde{\chi}^2 = 0.4$ and is true for any break mass between $2M_{\sun}$ and $3M_{\sun}$.  
The best fit broken power law is shown in Figure~\ref{difffig}.  
The best fit single power law ($\alpha=2.1\pm0.1$, $M>0.8~M_{\sun}$) is also shown, although this is not as good a fit to the data ($\tilde{\chi}^2 = 0.8$)

\citet{john00} report $\alpha_1 = 1.5$ ($M \lesssim 1 ~M_{\sun}$) and $\alpha_2 = 2.5$ ($M \gtrsim 1 ~M_{\sun}$) for $850~\micron$ sources in the central 700 arcmin$^2$ of Ophiuchus assuming a single dust temperature of 20~K and using Clumpfind photometry.  The Oph slopes are very close to those for the Perseus sample,
and if we assume the same dust temperature as \citet{john00} ($T=20$~K), the break mass of $\sim2.5~M_{\sun}$ becomes $\sim 1~M_{\sun}$ (the shape of the mass function does not change with $T_D$ as long as a single temperature is assumed).  Thus the submillimeter/millimeter mass distributions in Perseus and Ophiuchus are quite similar despite different environments, distances ($d=160$~pc for Ophiuchus, $d=250$~pc for Perseus), and resolutions ($14\arcsec$ vs. $31\arcsec$).
Comparisons are complicated, however, by blending, differing analysis and assumptions, and dust property uncertainties.

We also fit a lognormal distribution to the $M>0.8~M_{\sun}$ region where the mass function is reasonably complete:
\begin{equation}
\frac{dN}{dlogM} =  A exp \left[ \frac{-(logM-logM_0)^2}{2\sigma^2} \right],
\end{equation}
where
\begin{equation}
 \frac{dN}{dM} = \frac{1}{(ln10)M} \frac{dN}{dlogM}. 
\end{equation}
Here $A$ is a normalization factor, $\sigma$ is the width of the distribution, and $M_0$ is the characteristic mass.  The best-fitting lognormal for $M>0.8~M_{\sun}$ has $\sigma=0.5\pm0.1$ and $M_0=0.9\pm0.4~M_{\sun}$.  A lognormal is a somewhat better fit than a broken power law ($\tilde{\chi}^2 = 0.2$).  For $T_D=20$~K, the best fit parameters are $\sigma=0.5$ and $M_0=0.3~M_{\sun}$ (note again that only the characteristic mass, and not the shape, changes with $T_D$).  

Uncertainties in the dust temperature $T_D$ should have only a linear effect on the estimated mass as long as we are in the Rayleigh-Jeans (RJ) regime of the SED, where $B_{\nu}$ scales linearly with $T_D$.  At low temperatures, however, the departure of RJ from a true blackbody can cause large errors in the estimated mass.  If, for example, the true dust temperature of a source is $T_D=10$~K, assuming a temperature of $T_D=5$~K would cause a miscalculation of the mass by a factor of 4.7, and assuming a temperature of $T_D=20$~K would result in a factor of 3 error. 

\begin{figure}
%\plotone{f17.eps}
\plotone{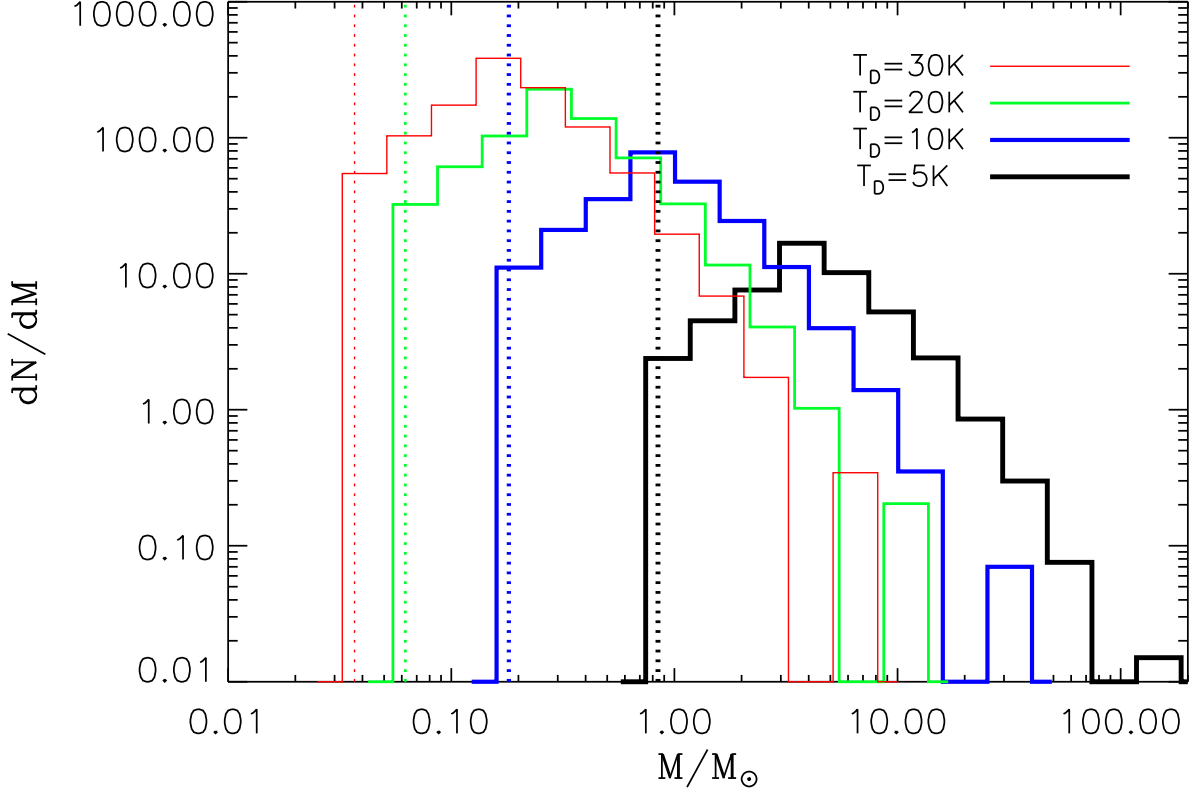}
\caption{The differential mass function, shown for different values of the assumed dust temperature $T_D$.  The dashed lines correspond to the mass limit for a point source at each $T_D$.  Changing the temperature shifts the distribution to higher (for lower $T_D$) or lower (for higher $T_D$) masses, but does not change its shape.  A dust opacity of $\kappa_{1.1mm} = 0.0114$~cm$^2$g$^{-1}$ and a distance of $d=250$~pc are assumed for all masses.  Increasing $\kappa_{1.1mm}$ shifts the distribution to lower masses, while increasing $d$ shifts it to higher masses.  Given the range of plausible values ($\kappa_{1.1mm}=0.005-0.02$~cm$^2$/g, $d=200-300$~pc), the effects of varying $\kappa_{1.1mm}$ and $d$ are smaller than the effect of varying the temperature ($T_D=5-30$~K).  \label{massfig}}
\end{figure}

Figure~\ref{massfig} demonstrates the effect on the differential mass function of varying the dust temperature.  
For each curve the mass function is calculated for a single value for $T_D$.  
Note that the shape of $dN/dM$ does not change with the assumed value of $T_D$, the distribution simply shifts to higher masses (lower $T_D$) or lower masses (higher $T_D$).  The shape \textit{will} change, however, if there is a range of dust temperatures in the real cores.  Ideally we could determine $T_D$ independently using observations at two different wavelengths.  
For our purposes $T_D =10$~K is a good compromise for prestellar and protostellar sources, but will overestimate by a factor of 3 the mass of sources with a true dust temperature of $20$~K.

Variations in $\kappa_{1.1mm}$ and $d$ have similar effects.
The dust opacity is uncertain by up to a factor of two or more \citep{oh94}, owing to large uncertainties in the assumed dust properties as well as the possibility that $\kappa_{\nu}$ varies with position within a core. 
Increasing the opacity shifts the mass distribution to lower masses, while increasing the distance shifts it to higher masses.  Both are smaller effects than changing the dust temperature for the range of plausible values ($\kappa_{1.1mm}=0.005-0.02$~cm$^2$~g$^{-1}$; $d=200-300$~pc; $T_D=5-30~K$).  
As for variable dust properties within the sample, blending of close sources will also distort the shape of the mass function, biasing it toward higher masses.  We know from previous observations that some sources are blends, but do not attempt to distinguish blends from single sources here. 
The total uncertainties in all masses are at least a factor of 4 or more \citep[e.g.][]{shir02,young03}.

 A comparison of the prestellar clump mass function to the stellar initial mass function (IMF) may reveal the origin of the IMF shape.  If stellar masses are determined by competitive accretion or by the protostars themselves through feedback mechanisms (e.g. outflows and winds), we would not expect the emergent IMF to reflect the original clump mass function \citep{AF96}.  If, on the other hand, stellar masses are determined by the initial fragmentation into cores, as might be expected in crowded regions where the mass reservoir is limited to a protostar's nascent core, the IMF should closely trace the clump MF \citep{myers98}.  
In Serpens \citep{TS98} and $\rho$ Oph \citep{motte98} the clump MF has been found to be quite similar to the stellar IMF, suggesting that fragmentation is responsible for determining final masses.
In reality, of course, it may well be a combination of these effects and turbulence that shapes the IMF \citep[e.g.][]{cb05}, but we consider the simplest cases here.

The local IMF follows a broken power law with $\alpha_1 = 1.6$ ($M<1~M_{\sun}$) and $\alpha_2 = 2.7$ ($M>1~M_{\sun}$), flattening around $0.3~M_{\sun}$ \citep{chab03}.  \citet{chab03} also find that the IMF is well fit by a lognormal with $\sigma=0.4$ and $M_0 = 0.5~M_{\sun}$.  Thus the slope of the IMF, but not the break mass, is very similar to the Perseus mass function.  Even the characteristic masses are quite similar if we assume $T_D=20$~K for the Perseus sample.  
Currently, a direct connection between the mass distribution in Perseus and the IMF is difficult to make because our sample contains sources at a range of ages, with varying amounts of the envelope already accreted onto the protostar or ejected in an outflow, and varying envelope temperatures. 
After combining these data with Spitzer c2d data it will be possible to determine the evolutionary state of each source, separating prestellar cores from more evolved objects.
Even for a sample containing only prestellar cores, however, the association with final stellar masses may be problematic \citep[see][]{john00}.

\subsection{Clustering}\label{cloudsect}

Visually, the 1.1~mm sources in Perseus appear very clustered: 89/122 or 73\%, have a neighboring source within $3 \arcmin$, with most isolated sources being faint objects near the detection limit.  In their $850~\micron$ SCUBA survey \citet{hatch05} find that 80\% of the $850 ~\micron$ sources are in groups of three or more sources (within 0.5~pc), and $40-60$\% of the sources are in the massive clusters of NGC~1333 and IC~348 (HH~211 in their paper).  Using similar criteria, we also find $\sim 80\%$ of the total number of sources lie in groups of $>3$ within $0.5$~pc, and $\sim 45-50\%$ in the massive clusters.  
Despite the fact that our 1.1~mm map covers more than twice as much area as the $850~\micron$ map of \citet{hatch05} ($7.5$~deg$^2$ compared to $\sim3$~deg$^2$), the clustering properties of sources in the two surveys are quite similar.  This result is perhaps not surprising considering that we only detect about $5-10$ sources in the additional $\sim 4.5$~deg$^2$ covered by the Bolocam map.  

For a more quantitative understanding of the clustering properties in Perseus, we calculate the two-point correlation function:
\begin{equation}
w(r) = \frac{H_s(r)}{H_r(r)} - 1,
\end{equation}
where $H_s(r)$ is the number of core pairs with separation between $log(r)$ and $log(r)+dlog(r)$.  The definition of $H_r(r)$ is similar to $H_s(r)$, but for a random distribution.  The random sample is constructed by generating a uniform random distribution of sources with the same RA and Dec limits as the real sample.  The two-point correlation function is often used in cosmological studies of the clustering of galaxies \citep[e.g][]{madd90}, but may also be a good way to compare the properties of different molecular clouds.  

\begin{figure}
\plotone{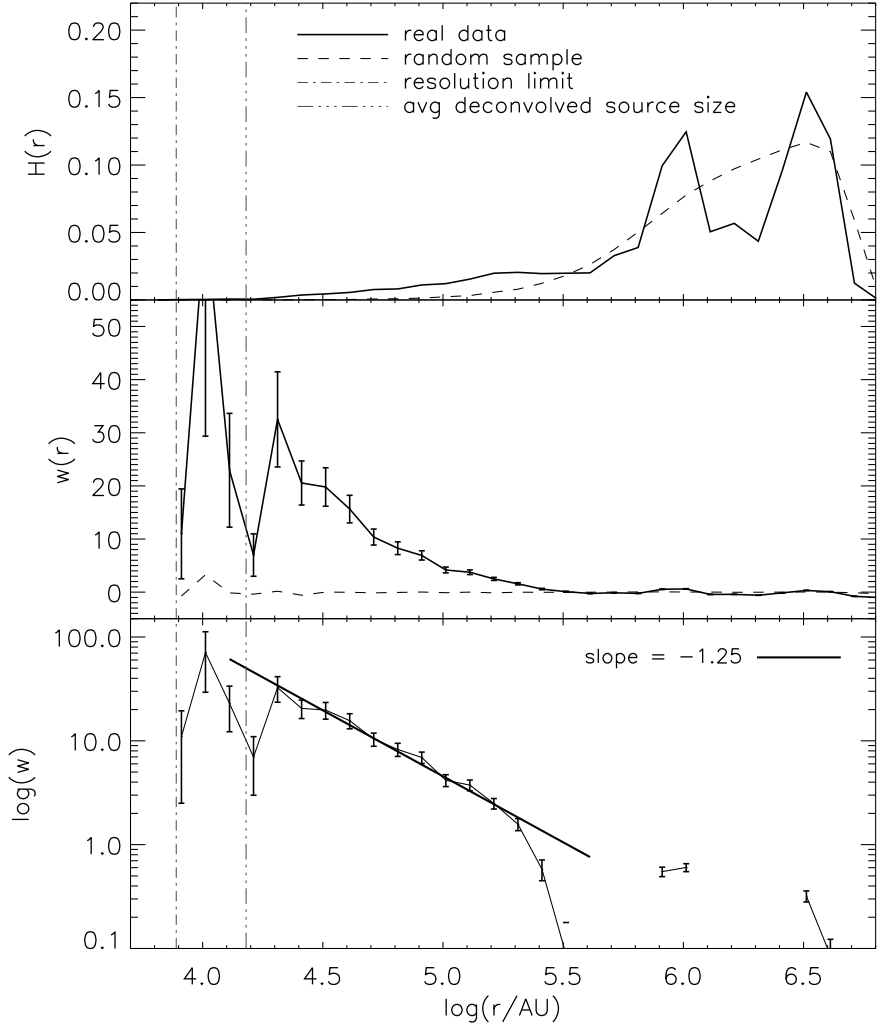}
\caption{The two-point correlation function, illustrating the degree of clustering in the cloud. Top: $H(r)$ is the number of source pairs between $log(r)$ and d$log(r)$ as a function of $log(r/AU)$.  The solid line indicates the real data, and the dashed line is for a uniform random distribution of sources with the same RA/Dec limits as the real sample. In all plots the dotted line denotes the resolution limit, and the dot-dash line the mean deconvolved source FWHM.  Center: The two point correlation function $w(r)$ as defined in the text, with $\sqrt{N}$ errors.  Where $w(r)>0$ there is a correlation between sources at that separation, thus $w(r)$ indicates clustering on scales $log(r/AU)< 5.5$ for the Perseus sample.  When calculated using two randomly distributed samples (dashed line) $w(r)$ shows no correlation, as expected. Bottom: $w(r)$ is well fitted by a power law, $w(r) \propto r^{-1.25}$, for $2.5 \times 10^4$~AU~$<r< 2 \times 10^5$~AU.  \label{corrfnfig}}
\end{figure}

Plots of $H_s(r)$ and $w(r)$ for the entire observed region of Perseus are shown in Figure~\ref{corrfnfig}, with the random distribution ($H_r(r)$, dashed line) included for comparison.  If the sources were randomly distributed in the cloud, we would expect the two curves to be similar.  The resolution limit ($31\arcsec=8\times10^3$~AU), and average deconvolved source FWHM size ($1.5\times10^4$~AU) are also indicated, the latter a being representation of the effective resolution limit.  It is clear that the source pair function $H_s(r)$ shows an excess over the random distribution $H_r(r)$ at small scales (the differences at large scales is not significant).  This is confirmed by the correlation function $w(r)$ (middle panel), which is $>3\sigma$ on scales $2\times10^4$~AU~$<r<2.5 \times 10^5$~AU.  Note that the random distribution shows no correlation ($w_r\sim0$), as expected.  

If we characterize the correlation function as a power law, $w(r) \propto r^{- \gamma}$, then a good fit ($\tilde{\chi}^2 = 0.7$) is obtained for $\gamma = 1.25\pm0.06$ in the range $2 \times 10^4$~AU~$<r< 2 \times 10^5$~AU.
Since the average deconvolved source size corresponds to $\sim 1.5\times 10^4$~AU, Perseus essentially shows clustering from the average source size up to $r_{max}=2\times 10^5$~AU~$=1.2$~pc.
In Ophiuchus \citet{john00} find $\gamma = 0.75$ for separations $r<3 \times 10^4$~AU and negligible clustering for $r>3\times 10^4$~AU, for a distance of 160~pc and a beam size of $2\times10^3$~AU.  Those authors associate $r_{max}=3\times10^4$~AU with the Jeans length in Ophiuchus, and note that $\gamma=0.7$ for galaxy clustering \citep{madd90}, where gravity is the dominant process.  It is possible that different slopes could be associated with different processes dominating fragmentation, but given the considerable uncertainties involved we choose not to speculate further.

While the correlation function provides a good comparison of clustering between the clouds, we note that the range over which $w(r)$ can be computed ($8 \times 10^3 - 5 \times 10^6$~AU for Perseus and $4 \times 10^3-10^5$~AU for Oph) differs due to the different resolution limits, spatial coverages, and distances of the two clouds.  Even so, it seems that clustering occurs on larger scales in Perseus than in Oph ($r_{max,Per}>r_{max,Oph}$), but drops off more quickly as a function of separation ($\gamma_{Per} > \gamma_{Oph}$). 
It is important to note that the measured correlation function depends on the area observed; if we break the Perseus map into smaller pieces the derived slope does vary, but it remains between $1<\gamma <1.5$ for a range of chosen areas.

\subsection{An extinction threshold for 1.1~mm cores}

The total mass contained in the 122 detected 1.1~mm cores is $285~M_{\sun}$, assuming $\kappa_{1.1mm}=0.0114$~cm$^2$~g$^{-1}$, $T_D=10$~K, and a gas to dust ratio of 100.  The total cloud mass based on CO observations is  $1-2\times10^4 ~M_{\sun}$ for $d=250$~pc \citep{sarg79,cern85,ut87,carp00}.  Based on the {\it{NICE}} extinction map, we calculated a total mass for $A_V \ge 2$ of $5900M_{\sun}$ in the area observed by Bolocam. 
Thus only a very small fraction, between 1\% and 5\%, of the cloud mass is contained in dense cores at $\lambda=1.1$mm.  This small fraction is consistent with the evidence from comparisons of molecular cloud masses to total stellar masses that molecular cloud material is relatively sterile \citep{evans99}.  \citet{hatch05} find a significantly larger fraction of mass (20\%) in $850~\micron$ emission.  Their large total mass at $850~\micron$ of $\sim2600~M_{\sun}$ is due in part to differences in the assumed distance and opacity, and in part to the fact that those authors integrate over all $>5\sigma$ pixels in the map to get the total mass, whereas we only include only the mass in discrete cores.

The low efficiency of 1.1~mm cores may be related to the conditions required for a dense core to form.  
\citet{john04} have recently suggested that there is an extinction threshold for forming $850~\micron$ cores in Ophiuchus.  
The derived threshold for the presence of stable cores in Ophiuchus ($A_{V,lim}^{Oph}\sim15$~mag) is greater than the maximum extinction in Perseus as traced by our ${\it{NICE}}$ extinction map ($A_{V,max}^{Per}\sim16$~mag), suggesting that if conditions in the two regions were similar there should be very few millimeter cores in Perseus, which is clearly not the case.  
In fact, the mean extinction from our ${\it{NICE}}$ map toward all cores in Perseus is $\mean{A_V}=7.1\pm3.1$~mag, similar to the minimum $A_V$ at which any $850~\micron$ core is found in Ophiuchus ($A_{V,min}^{Oph}=7$~mag; \citealt{john04}). 
Bear in mind, however, if the cloud material is very clumpy, beam effects will be important given that the distance to Perseus is twice that of Ophiuchus.
For comparison, the mean $A_V$ for the entire area observed by Bolocam is $\mean{A_V}\sim2$~mag, and the mean $A_V$ for the area observed by \citet{john04} in Ophiuchus was $\mean{A_V}\sim4$~mag.  

\begin{figure}
\plotone{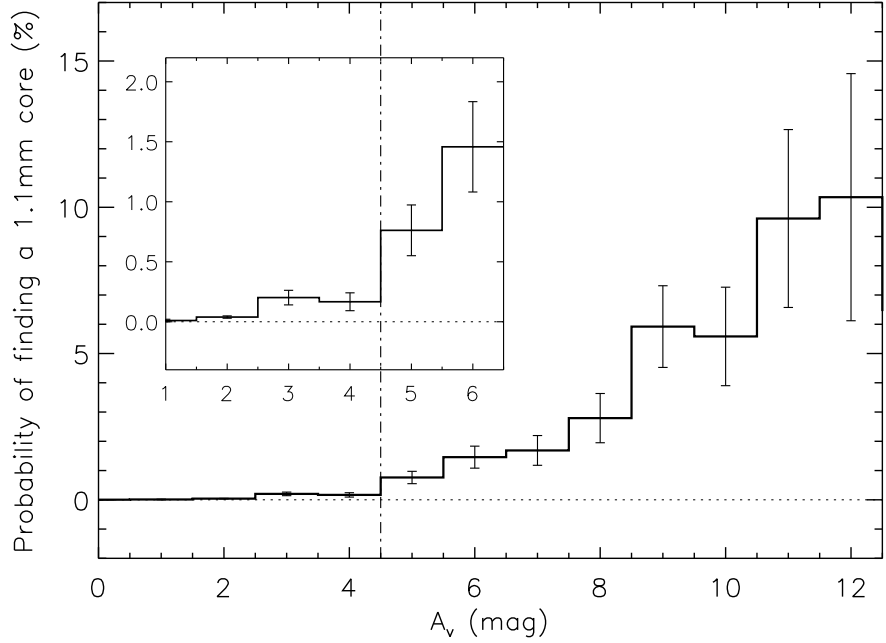}
\caption{Probability of finding a 1.1~mm core as a function of $A_V$, computed using the ${\it{NICE}}$ extinction map.  The probability is the number of $50\arcsec$ pixels at a given $A_V$ containing one or more 1.1~mm cores, divided by the total number of pixels at that $A_V$. Error bars are Poisson statistical errors, and become large where the $A_V$ map saturates at $A_V\sim10$~mag.  The low $A_V$ region is magnified for clarity.  It appears that there is an approximate extinction limit at $A_V\sim5$~mag, below which it becomes very unlikely that a 1.1~mm core will be found.   The extinction limit of $A_V\sim 5$~mag is considerably lower than the $A_V\sim 15$~mag limit for forming $850~\micron$ cores found by \citet{john04} in Ophiuchus, although beam effects could be important.  Our $A_V\sim 5$~mag limit is consistent with the fact that few sources are found outside previously surveyed regions.  
\label{avprobfig}}
\end{figure}

All dense groups of 1.1~mm sources in Perseus lie in regions of high $A_V\gtrsim 5$~mag, but not all regions of high $A_V$ correspond to 1.1~mm sources, suggesting that relatively high extinction may be necessary, but not sufficient, for star formation.  
The probability of finding a 1.1~mm core as a function of $A_V$ is shown in Figure~\ref{avprobfig}.  The probability for a given $A_V$ is calculated from the extinction map as the number of $50\arcsec$ pixels containing a 1.1~mm core divided by the total number of pixels at that $A_V$ ($p/100=N_{src}/N_{A_V}$).  Error bars are Poisson statistical errors ($\sigma_p/100 = \sqrt{N_{src}}/N_{A_V}$).  
It appears that there is an approximate extinction limit of $A_V\sim5$~mag below which it becomes very unlikely that a 1.1~mm core will be found.  Above $A_V\sim5$~mag, the probability of finding a 1.1~mm core rises with $A_V$.  Error bars are large for high $A_V$ because our ${\it{NICE}}$ $A_V$ map is not sensitive to $A_V\gtrsim10$~mag, so there are few pixels at high extinctions.

\citet{hatch05} take a more sophisticated approach to the probability of finding $850~\micron$ cores as a function of column density, concluding that there is no column density (or $A_V$) limit for submillimeter cores in Perseus.  
This result is consistent with our extinction limit of $A_V\sim 5$~mag given that the \citet{hatch05} survey covers only the $A_V\gtrsim4$~mag regions in Perseus.  An extinction threshold around $A_V\sim5$~mag in Perseus would also explain why very few ($5-10$) sources are found in the Bolocam 1.1~mm map outside of the region covered by the \citet{hatch05} survey, despite the much greater area imaged.  The additional area covered is primarily low column density $A_V\sim2-4$~mag material, and therefore would not be expected to contain many millimeter cores.

\subsection{Comparison to c2d Observations:  B1 Ridge}\label{b1sect}

\begin{figure}
\epsscale{1.1}
\plottwo{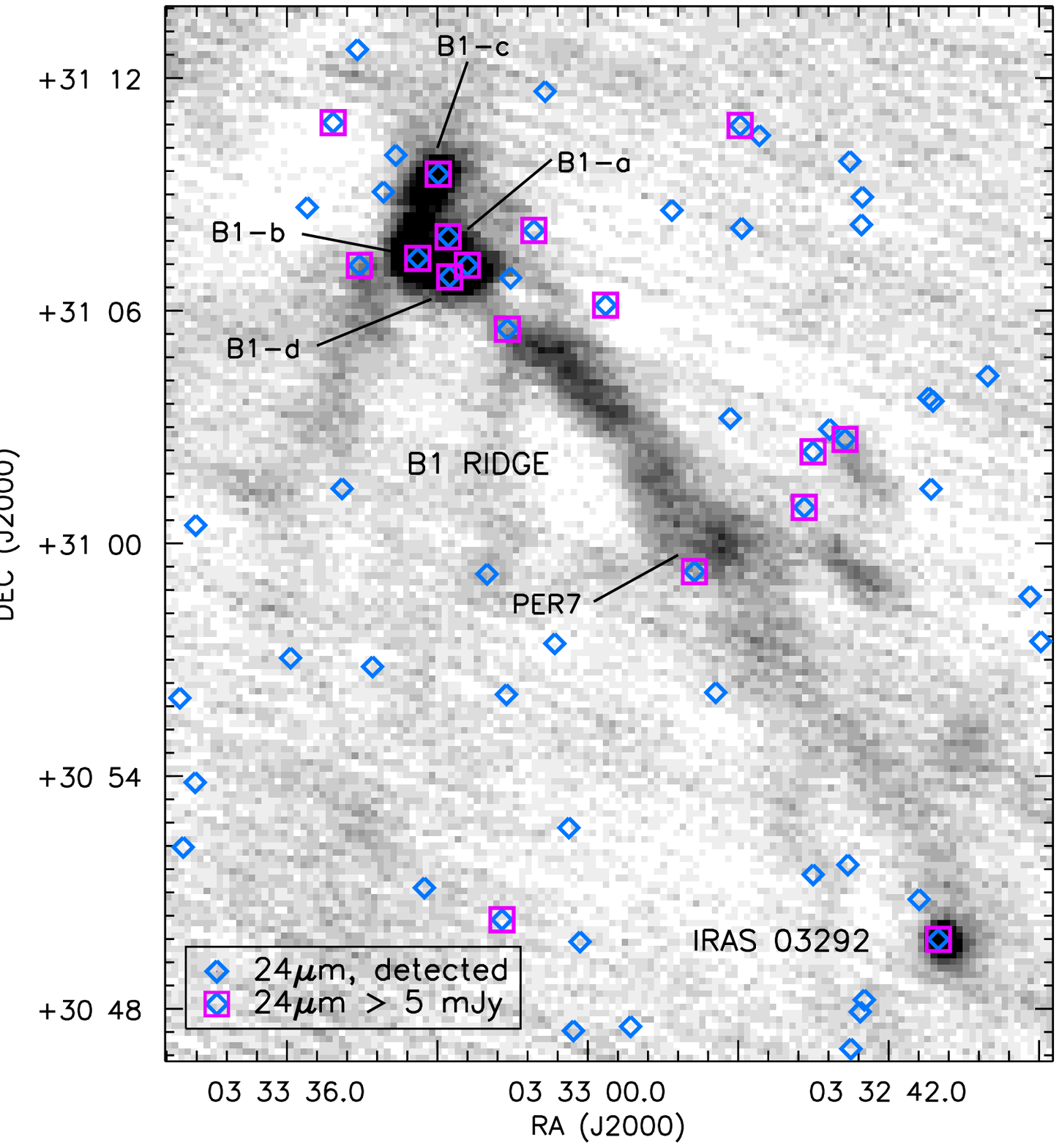}{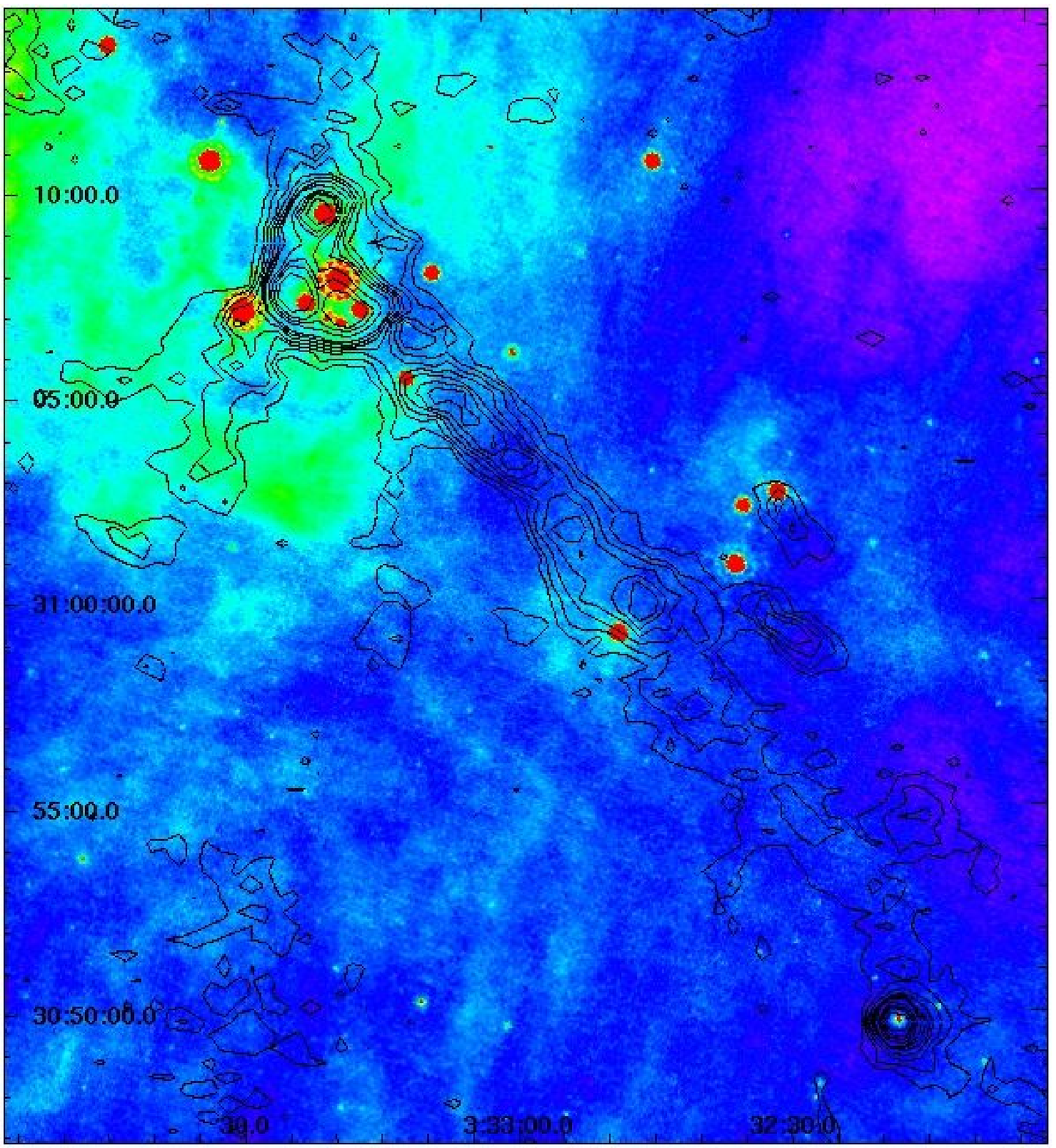}
\caption{Spitzer sources in the B1 Ridge region overlaid on the Bolocam 1.1~mm map (left).  Blue diamonds show the positions of all detected $24~\micron$ sources in the MIPS c2d image (Rebull et al. 2005, in preparation), and pink boxes indicate sources with $S_{24}>5$~mJy.  The MIPS $24~\micron$ image is also shown (right) with Bolocam 1.1~mm contours overlaid ($2,4,..20\sigma$). There are no $24~\micron$ sources in the main part of the B1 Ridge, suggesting that the ridge may be made up of a number of prestellar cores.  \label{overlayfig}}
\epsscale{1}
\end{figure}

As an example of the analysis of the complementary \textbf{c2d} Spitzer IR and Bolocam 1.1~mm observations, we compare the Spitzer $24~\micron$ and 1.1~mm images for a small area of Perseus around B1.  The B1 Ridge is a narrow ridge of extended 1.1~mm emission below the group of protostars B1a-d.  
Figure~\ref{overlayfig} shows the B1 Ridge region of the Bolocam map (left), with the position of all detected MIPS 24$~\micron$ sources, as well as MIPS sources with $S_{24}>5$~mJy, indicated.  The c2d MIPS $24~\micron$ image is also shown (right) with Bolocam 1.1~mm contours overlaid.
The B1(a-d) protostellar sources are bright at both $24~\micron$ and 1.1~mm, with $(S_{24},S_{1.1mm}) = (0.2,1.2)$~Jy/beam (B1-b), $(0.8,1.1)$~Jy/beam (B1-d) and $(0.13,0.6)$~Jy/beam (B1-d).  The little-known protostar IRAS 03292+3039 is relatively faint at $24~\micron$ ($0.08$~Jy/beam), but very bright at 1.1~mm (1.1~Jy/beam), suggesting a young evolutionary state.  

There are a few bright $24\micron$ protostellar sources nearby and in the lower part of the extended B1 ridge (e.g. near the ammonia core Per~7 \citep{ladd94}), but in the main part of the ridge there are no $24~\micron$ sources.  
Given the lack of mid-IR sources in the main B1 ridge, we suggest that it is made up of a number of prestellar cores.  High resolution submillimeter and millimeter follow-up observations of both the ridge and nearby protostars have already been completed with OVRO ($\lambda=3$mm) and SHARC II ($\lambda=350~\micron$).  
A detailed paper focusing on this region that will combine IRAC and MIPS data with  high resolution millimeter data to study the precise nature of the prestellar cores and embedded protostars, including envelope structure and outflow dynamics, is in preparation.

\section{Summary}\label{sumsect}

We present a 7.5 deg$^2$ (143 pc$^2$ for $d=250$~pc) survey for 1.1~mm dust continuum emission in the Perseus molecular cloud using Bolocam at the CSO.  
This map is the largest millimeter or submillimeter continuum image of the region to date.  
Given that Bolocam has a beam of $31\arcsec$ (FWHM), the Perseus map covers a remarkable total number ($3.4\times 10^4$) of resolution elements down to a $5\sigma$ point source mass detection limit of $0.18~M_{\sun}$ ($T_D=10$~K). 

We detect 122 1.1~mm cores above the $5\sigma=75$mJy/beam detection limit. 
Nearly half (60/122) of the detected sources are new millimeter or submillimeter detections, including previously unknown sources as well as known objects not previously observed at these wavelengths.
Our 1.1~mm map covers more than twice the area of the recent $850~\micron$ SCUBA survey by \citet{hatch05} but, despite the significantly greater area imaged, only $\sim5-10$ of our new detections lie outside their map.  Thus, much of the Perseus cloud is devoid of compact millimeter emission and, by implication, active star formation. 
%The remaining $\sim50$ new sources, which are generally faint and near known groups, are detected as a result of the lower mass limit of this survey ($\sim0.2~M_{\sun}$) compared to the SCUBA survey ($\sim 0.4~M_{\sun}$).

In order to compare the general morphology of our 1.1~mm map to the COMPLETE $^{13}$CO map and our ${\it{NICE}}$ extinction map, we convert all three images to a column density scale. 
Our 1.1~mm map reveals significantly higher column density features than the other tracers and exhibits much more compact structure, even when degraded to the $5\arcmin$ resolution of the ${\it{NICE}}$ extinction map.  The general appearance of the 1.1~mm emission is roughly consistent with the molecular and extinction data, however, in that most 1.1~mm sources lie within $^{13}$CO and $A_V$ peaks.

The total mass in discrete 1.1~mm cores is $285~M_{\sun}$ ($T_D=10$~K), accounting for no more than 5\% of the total mass of the cloud.  The small fraction of mass in dense cores, which are usually associated with star formation, supports the idea that most of the mass in molecular clouds is relatively sterile \citep{evans99}.  
Calculating the probability of finding a 1.1~mm core as a function of $A_V$ leads us to conclude that there is an extinction threshold in Perseus at $A_V\sim5$~mag, above which 1.1~mm cores are likely to be observed.  Such an extinction limit is consistent with the fact that very few new sources are found outside of the area covered by the $A_V\gtrsim4$~mag map of \citet{hatch05}.

The average mass of the sample, based on the total flux density from aperture photometry, is $2.3 ~M_{\sun}$ ($T_D=10$~K). 
The differential mass function $dN/dM$ is well fitted by a broken power law with $\alpha\sim1.3$ ($0.5~M_{\sun}<M<2.5M_{\sun}$) and $\alpha\sim2.6$ ($M>2.5~M_{\sun}$). 
The derived values are similar to those found in Ophiuchus ($\alpha = 2.5 ~M > 1 ~M_{\sun}, \alpha=1.5~ M<1~M_{\sun}$ \citealt{john00}), and to the local IMF ($\alpha = 2.7~ M>1~M_{\sun}, \alpha = 1.6~ M<1~M_{\sun}$, \citealt{chab03}). 
We compute the two-point correlation function, confirming that compact millimeter emission in the cloud is highly structured.  Significant clustering of sources from the average source size up to scales of $2\times10^5$~AU is seen.  Within this range, the correlation function exhibits a power law shape with index $\gamma=-1.25$, steeper than the correlation in Ophiuchus \citep{john00}.

The Bolocam 1.1~mm data presented here were designed to cover the same region as the Spitzer c2d legacy program IRAC ($\lambda=3.6-8.0 ~\micron$) and MIPS ($\lambda=24-160 ~\micron$) maps of Perseus.
Combining the 1.1~mm data with the c2d IRAC and MIPS data will enable a complete census of the properties and distribution of the protostars and dense cores in Perseus, especially when complemented by the publicly available 2MASS catalogs and COMPLETE molecular and continuum maps. 
Follow-up observations have already been completed with SHARC II (CSO, $\lambda=~\micron$) and the OVRO interferometer ($\lambda=3$mm) for several of the most interesting sources identified in the Bolocam map, including the B1 ridge.  A detailed discussion of these regions utilizing Spitzer photometry, high resolution submillimeter and millimeter images, molecular line data, and SED and radial profile modeling is in preparation.

\acknowledgments

We gratefully thank the referee, Alyssa Goodman, as well as Karl Stapelfeldt and Philip Myers, for their helpful comments.
Support for this work, part of the Spitzer Legacy Science
Program, was provided by NASA through contracts 1224608 and 1230782
issued by the Jet Propulsion Laboratory, California Institute of
Technology, under NASA contract 1407.  Additional support was obtained
from NASA Origins Grant NNG04GG24G to the University of Texas at Austin.
Support for the development of Bolocam was provided by NSF grants AST-9980846 and AST-0206158.
MLE acknowledges support of an NSF Graduate Research Fellowship.

\clearpage

%% the file, uncomment the \plotone, \plottwo, and
%% If you need a layout that cannot be achieved with \plotone or
%% \plottwo, you can invoke the graphicx package directly with the
%% \includegraphics command or use \plotfiddle. For more information,
%% please see the tutorial on "Using Electronic Art with AASTeX" in the
%% documentation section at the AASTeX Web site,
%% http://www.journals.uchicago.edu/AAS/AASTeX.
%%
%% The examples below also include sample markup for submission of
%% supplemental electronic materials. As always, be sure to check
%% the instructions to authors for the journal you are submitting to
%% for specific submissions guidelines as they vary from

%% This example uses \plotone to include an EPS file scaled to
%% 80% of its natural size with \epsscale. Its caption
%% has been written to indicate that additional figure parts will be
%% available in the electronic journal.

\notetoeditor{Figure 2 should be color in the printed version.  Please convert to CMYK for the printed version only, and try to match the colors of the submitted printed image.}

\notetoeditor{Figure 3 should be online-only color.}

\notetoeditor{Figure 4 should be online-only color.}

\epsscale{0.7}
\notetoeditor{Figure 5 should be full page.}

\epsscale{0.9}
\notetoeditor{Figure 6 should be full page, and color in the printed version.}

\epsscale{0.85}
\notetoeditor{Figure 7 should be full page if possible, and color in the printed version.}
\epsscale{1.0}

\clearpage

\notetoeditor{Figures 8,9,10 should be placed together if possible.}
\epsscale{0.7}

\epsscale{1.0}

\notetoeditor{Figure 11 should be online-only color.}

\notetoeditor{Figure 12 should be online-only color.}

\notetoeditor{Figure 16 should be online-only color.}

\notetoeditor{Figure 17 should be online-only color.}

\epsscale{0.75}
\epsscale{0.8}

\notetoeditor{Figures 20a and 20b should be color in the printed version.  Please try to have the body of the figures (without axes labels) the same size.  Please convert to cmyk for the printed version only.}

\notetoeditor{Please convert tables to machine readable format.}
\input{tab1}
\notetoeditor{Table 2 should be online only, with a sample table appearing in the printed version.  tab2.tex is the full table, tab2.sample.tex is the sample to appear in the printed version.}
\input{tab2}

%% The following command ends your manuscript. LaTeX will ignore any text

\end{document}